\documentclass[acmsmall]{acmart}

\usepackage{algorithmic}
\usepackage{graphicx}
\usepackage{textcomp}
\usepackage{xcolor}
\usepackage{comment}
\usepackage{hyperref}
\usepackage{bm}
\usepackage{enumitem}
\usepackage{appendix}

\usepackage{amsmath}
\usepackage{lipsum}
\usepackage{multirow}
\usepackage[justification=centering]{caption}
\usepackage{caption}
\usepackage{url}
\usepackage{listings}
\usepackage{subfigure}
\usepackage{array}
\usepackage{pifont}

\setcopyright{cc}
\setcctype{by}
\acmDOI{10.1145/3729347}
\acmYear{2025}
\acmJournal{PACMSE}
\acmVolume{2}
\acmNumber{FSE}
\acmArticle{FSE077}
\acmMonth{7}
\received{2024-09-12}
\received[accepted]{2025-04-01}

\newcommand{\changed}[1]{{#1}}  

\begin{document}

\title{Incorporating Verification Standards for Security Requirements Generation from Functional Specifications}

\author{Xiaoli Lian}
\affiliation{
  \institution{SKLCCSE, Beihang University}
  \city{Beijing}
  \country{China}
}
\email{lianxiaoli@buaa.edu.cn}

\author{Shuaisong Wang}
\affiliation{
  \institution{Beihang University}
  \city{Beijing}
  \country{China}
}
\email{littletree@buaa.edu.cn}

\author{Hanyu Zou}
\affiliation{
  \institution{Beihang University}
  \city{Beijing}
  \country{China}
}
\email{zouhanyu@buaa.edu.cn}

\author{Fang Liu}
\affiliation{
  \institution{Beihang University}
  \city{Beijing}
  \country{China}
}
\email{fangliu@buaa.edu.cn}

\author{Jiajun Wu}
\affiliation{
  \institution{Beihang University}
  \city{Beijing}
  \country{China}
}
\email{22373141@buaa.edu.cn}

\author{Li Zhang}
\authornote{This author is the corresponding author.}
\affiliation{
  \institution{SKLCCSE, Beihang University}
  \city{Beijing}
  \country{China}
}
\email{lily@buaa.edu.cn}

\begin{abstract}
In the current software-driven era, ensuring privacy and security is critical. Despite this, the specification of security requirements for software is still largely a manual and labor-intensive process. Engineers are tasked with analyzing potential security threats based on functional requirements (FRs), a procedure prone to omissions and errors due to the expertise gap between cybersecurity experts and software engineers. To bridge this gap, we introduce \textit{F2SRD} (Function-to-Security Requirement\changed{s} Derivation), an automated approach that proactively derives security requirements (SRs) from functional specifications under the guidance of relevant security verification requirements (VRs) drawn from the well recognized OWASP Application Security Verification Standard (ASVS).
\textit{F2SRD} operates in two main phases: Initially, we develop a VR retriever trained on a custom database of FR-VR pairs, enabling it to adeptly select applicable VRs from ASVS. This targeted retrieval informs the precise and actionable formulation of SRs. Subsequently, these VRs are used to construct structured prompts that direct GPT-4 in generating SRs. Our comparative analysis against two established models demonstrates \textit{F2SRD}'s enhanced performance in producing SRs that excel in \emph{inspiration, diversity, and specificity}—essential attributes for effective security requirement generation. By leveraging security verification standards, we believe that the generated SRs are not only more focused but also resonate stronger with the needs of engineers.
  
\end{abstract}

\ccsdesc[500]{Software and its engineering~Requirements analysis}

\keywords{Security Requirements, Security Verification Standards, Functional Requirements}

\maketitle

\section{Introduction}
\label{sec:intro}

Today, a multitude of software applications permeate every aspect of our lives, serving as repositories for critical data—including banking and health information—and monitoring our daily activities. In this context, the security attributes of software become increasingly vital, as they determine the software's ability to withstand and counteract security threats effectively.  The alarming statistics from Contrast Security’s 2020 Application Security Observation Report \cite{contrastSecurity} reveal that 96\% of web applications harbor vulnerabilities, with 26\% facing severe issues. Furthermore, reports by CISQ—the Consortium for Information and Software Quality—underscore an escalating trend in cyberattacks and their associated costs to the global economy \cite{CPSQReport}. Against such a backdrop, ensuring software security has become a critical concern, commanding attention in both academia and industry \cite{ANSARI2022191, inbook}.

\changed{Security requirements (SRs) specify the necessary protections for software or systems against threats such as unauthorized access, data breaches, and various forms of cyber-attacks.} Establishing clear, comprehensive, and effective SRs at the onset of a software project is essential for ensuring robust security \cite{ANSARI2022191}. These SRs are instrumental in guiding the design, development, and security testing phases, ultimately bolstering the security of the final software product.

In practice, however, most software projects tend to prioritize functional requirements \changed{(FRs, also referred to as functional specifications) — which detail what the system is expected to do} — over SRs \cite{9397153, 10.1145/1082983.1083214}. While engineers diligently analyze FRs, SRs are often relegated to a secondary role, sometimes only considered after the main \changed{FRs} have been established or, worse yet, deferred until the testing phase \cite{9736282}. This oversight primarily manifests in one pervasive issue: \emph{the failure to adequately capture essential SRs}. This is often due to two contributing factors. Firstly, SRs are typically addressed in isolation from the broader requirements engineering process, which can lead to critical security features being overlooked. Secondly, a knowledge gap exists between requirements engineers, who may lack comprehensive cybersecurity expertise, and security specialists, who might not be thoroughly acquainted with the system's details, further amplifying the likelihood of missing vital SRs \cite{10.1007/978-3-642-19858-8_2, 10.1007/s10664-016-9451-7, 6912260}. Unfortunately, the prevalence of incomplete and/or hidden requirements stands out as the most commonly reported challenge according to data from 228 companies across ten countries and various industries \cite{10.1007/s10664-016-9451-7}. Therefore, \emph{the automatic generation of SRs, with the guidance of cybersecurity domain knowledge, as soon as the FRs are defined becomes imperative}.

Recent studies on the (semi-)automation of SRs generation have largely been categorized into two streams: traditional (non-deep learning) approaches and deep learning-based methods. Traditional strategies, like model-based \cite{10.1007/s00766-017-0279-5, turetken2004automating} and template-based generation \cite{6912260}, are typically resource-intensive and costly. These approaches usually require detailed preparatory efforts, such as crafting domain ontologies \cite{10.1007/s00766-017-0279-5}, accurate structural models \cite{turetken2004automating, DBLP:journals/infsof/CoxPBV05, DBLP:journals/re/MaidenMJG05, DBLP:conf/coopis/YuBDM95, DBLP:conf/sigsoft/LetierL02, DBLP:journals/re/LandtsheerLL04, van2004goal, DBLP:journals/tse/LamsweerdeW98, DBLP:journals/re/MezianeAA08, DBLP:conf/re/Berenbach03a,9793770} 
or pre-established rules \cite{6912260} — steps known for being time-consuming. 
On the other hand, over the past couple of years, deep learning-based methods for SR synthesis have started to emerge. For example, Koscinski et al. \cite{koscinski2023demand} recommended employing RelGAN \cite{nie2018relgan} to synthesize SRs. However, given RelGAN's goal to create realistic requirements using its training data—which contains both functional and security requirements across various projects —there's a significant likelihood that its outputs may not qualify as SRs, may be irrelevant to specific projects, or could contain syntactical inaccuracies \cite{koscinski2023demand}.

\begin{figure}[!htbp]
	\centering
	\includegraphics[trim = {0 4.6cm 0cm 0}, clip, width=0.95\textwidth]{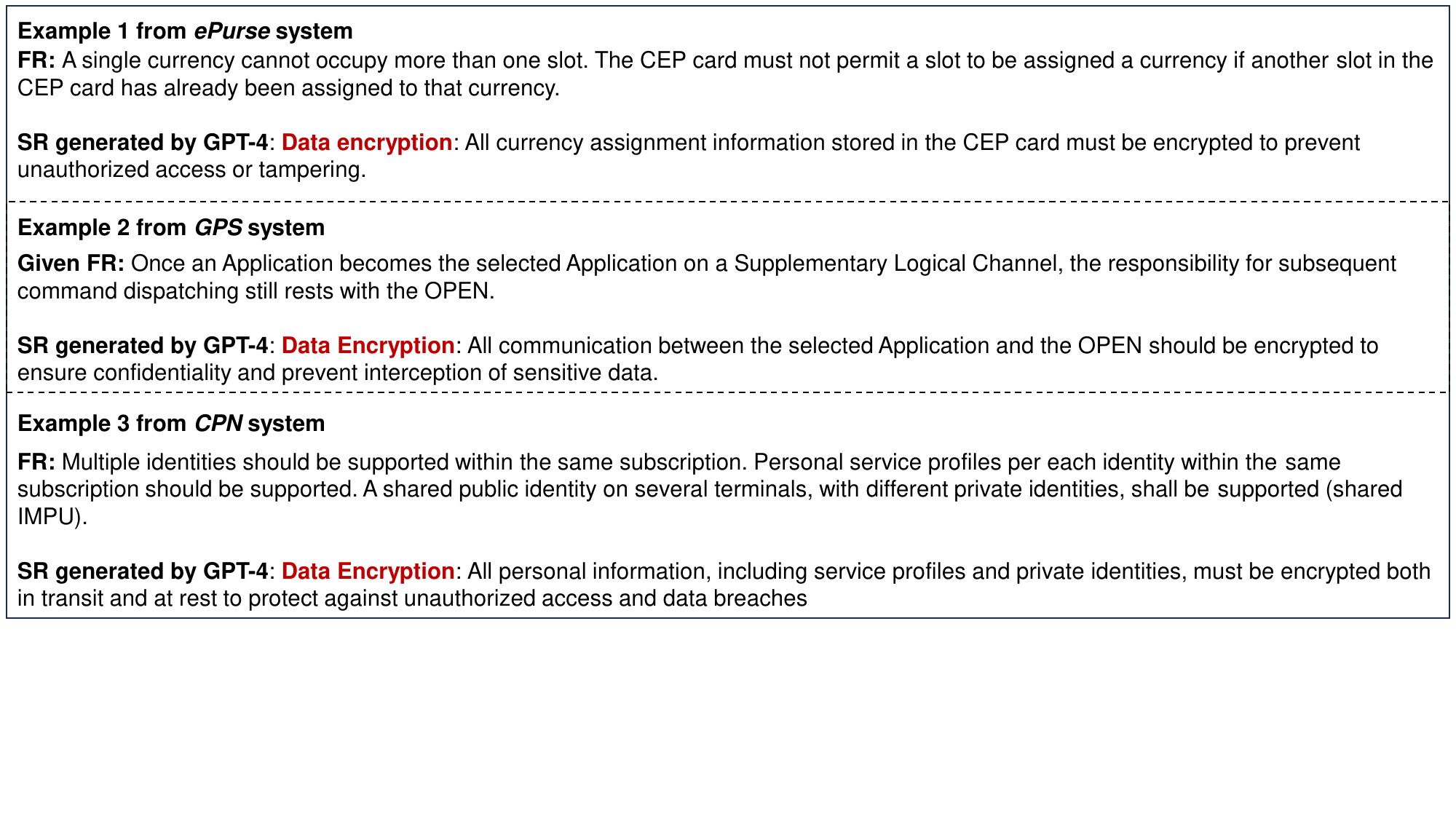}
	\caption{Examples of GPT-4 Generating Generic Security Requirements for Diverse Functional Requirements.}
	\label{fig:GPTExample}
\end{figure}

Advanced Large Language Models (LLMs) like GPT-4 have shown impressive aptitude in interpreting and generating human-like text \cite{achiam2023gpt, dubey2024llama}. While GPT-4 can generate contextually appropriate, syntactically correct recommendations aligned with specified FRs, our research reveals a shortfall: without detailed, specific input, the model tends to produce generic, well-known SRs that offer limited utility for engineers. As demonstrated by three cases in Figure \ref{fig:GPTExample}, when tasked with diverse FRs from different software systems, GPT-4 regularly recommends data encryption. Though technically sound, this advice lacks the specificity needed for practical security enhancements in unique software environments. For example, in the first case study, considerations such as slot assignment verification and encryption key management are also critical for robust security.

The challenge for GPT-4 to generate sound SRs is twofold: 1) defining relevant SRs guidance, and 2) sourcing this information. In practice, engineers often fortify and guarantee the security of specified FRs by consulting established security verification standards. These standards articulate detailed verification requirements (VRs) that specify the various aspects necessary for securing a system. Additionally, a practical advantage of aligning SRs with these standards is the ability to \changed{``standardize the process''}, thereby ensuring that the developed system achieves compliance with standardized verification protocols more effectively.

Inspired by this insight, our methodology integrates pertinent VRs into the GPT-4 prompting instructions. We have chosen the OWASP Application Security Verification Standard (ASVS) version 4.0.3 \cite{ASVS}, which comprises 286 VRs designed to secure systems. The ASVS is acknowledged as an authoritative global industry benchmark for application security and is embraced by a host of organizations (e.g., Booz Allen Hamilton\footnote{https://www.boozallen.com/} and CGI Federal\footnote{https://www.cgi.com/us/en-us/federal}), academic researchers \cite{8860028, 9255559}, and educators \cite{9402211}. 

We introduce \textit{F2SRD} (Functional-to-Security Requirements Derivation), an approach designed to \emph{automatically generate SRs tightly aligned with security standards and specific FRs contexts, providing clear, actionable guidance to reinforce the software system's functional security.}
This method unfolds over two phases: the design and training of a VR retriever, followed by the generation of SRs. To pinpoint VRs corresponding to FRs, we crafted a VR retriever inspired by the ColBERTv2 architecture \cite{Santhanam2021ColBERTv2EA}—a state-of-the-art information retrieval model. Given the scarcity of existing FR-VR pair data, we generated a synthetic dataset by creating ten FRs for each VR. After refining this dataset through relevance filtering, we trained our retriever with 1536 vetted FR-VR pairs. In the SR\changed{s} generation phase, the retriever suggests five VRs relevant to the input FRs. We then leverage GPT-4, guiding it to generate SRs by considering the relevancy between VRs and FRs, focusing on those VRs essential for verification against the submitted FRs. 

In our experiments, we initially assessed the performance of our VR retriever, which achieved an average accuracy rate of 80.4\%. In comparison, the accuracy rate for human evaluators identifying VRs was approximately \changed{85.1}\%. However, we observed only moderate agreement among different individuals regarding VR recommendations, highlighting the challenges of human identification and underscoring the practical value of our retrieval system.
For the SRs generated by our approach, we prioritized three key attributes: inspiration, diversity, and specificity. We conducted benchmark tests against RelGAN \cite{koscinski2023demand} and GPT-4. The outcomes from both quantitative and qualitative assessments indicate that our  \textit{F2SRD}, significantly surpasses these baselines. Additionally, we discuss the failure scenarios to inspire future research.


The contributions of our work are threefold:

\begin{itemize}[leftmargin=0.4cm]
    \item To the best of our knowledge, this is the first work to integrate security verification requirements from widely-applied standards with LLMs for the automatic synthesis of SRs.
    
    \item Both qualitative and quantitative analyses demonstrate the effectiveness of our approach. Furthermore, the evaluation framework we developed offers a reusable methodology for future research in this domain.
    
    \item The establishment of a new baseline for SR\changed{s} generation, with resources and datasets made available to the public via figshare.
\end{itemize}

\section{Background: OWASP Application Security Verification Standard (ASVS) and Motivating Example}
\label{sec:background}

\subsection{OWASP ASVS}
The OWASP Application Security Verification Standard (ASVS) is a flagship initiative of the Open Web Application Security Project (OWASP), a non-profit organization committed to enhancing software security\footnote{\url{https://owasp.org/www-project-application-security-verification-standard/}}. Garnering substantial recognition with over 2.6k stars on GitHub (till 2024-07-30), ASVS has been a cornerstone project since its inception in 2008, aiming to elevate the security of software around the globe, with a particular focus on web and API-based applications.

ASVS plays a critical role in establishing trust in the security robustness of web applications and has emerged as a de facto global industry standard for application security. Numerous organizations, such as Booz Allen Hamilton and CGI Federal, renowned for their expertise in security consulting and IT services, have adopted ASVS into their security protocols, underscoring its thoroughness and reliability in addressing application security challenges \cite{ASVS}. Beyond its industrial uptake, several academic researchers have proposed various methods for augmenting system security using ASVS as a foundation \cite{8860028, 9255559}. Additionally, educators are advocating for the integration of ASVS into Software Security Courses to enhance educational standards \cite{9402211}.

\changed{We use the latest version 4.0.3, which contains a comprehensive list of 277 valid VRs (out of a total of 286, with 9 deprecated).} These VRs have been collaboratively developed and ratified by a broad security community. \emph{They delineate the security verification prerequisites essential for the design, development, and testing of contemporary secure applications.}
For enhanced clarity, we detail the contents of these VRs in Table \ref{tab:ASVS}, showcasing the topics, the quantity of VRs per topic, and an exemplary VR for each category. It is evident that the VRs are systematically organized into 14 key aspects. Beyond conventional security protocols—authentication, session management, and access control, as referenced in \cite{SoftwareArchitecture}—this version also extends to cover early architectural considerations, defense against malicious code, file and resource management, and system configuration, thereby encompassing the full spectrum of application security concerns.

  \begin{table*}[!htbp]
    \centering
   \small
    \caption{The Information of Verification Requirements (VRs) in ASVS.}
    \label{tab:ASVS}
        \begin{tabular}{|c|p{2cm}|c|p{9cm}|} \hline
           \textbf{ID} & \textbf{Chapter}  & \textbf{\#VRs} & \textbf{Example} \\ \hline
           V1 &  Architecture, design and threat modeling & \changed{39} & 1.1.1 Verify the use of a secure software development lifecycle that addresses security in all stages of development. \\ \hline
           V2& Authentication & 57  & 2.1.1 Verify that user set passwords are at least 12 characters in length (after multiple spaces are combined). \\ \hline
           V3 & Session management & 20 & 3.1.1 Verify the application never reveals session tokens in URL parameters. \\ \hline
           V4 & Access control & \changed{9} &4.1.2 Verify that all user and data attributes and policy information used by access
controls cannot be manipulated by end users unless specifically authorized. \\ \hline
V5& Validation, sanitization and encoding & 30 & 5.1.3 Verify that all input (HTML form fields, REST requests, URL parameters, HTTP headers, cookies, batch files, RSS feeds, etc) is validated using positive validation (allow lists). \\ \hline
V6 & Stored cryptography & 16  & 6.1.2 Verify that regulated health data is stored encrypted while at rest, such as
medical records, medical device details, or deanonymized research records. \\ \hline
V7&Error handling and logging & \changed{12} &7.1.2 Verify that the application does not log other sensitive data as defined under
local privacy laws or relevant security policy. \\  \hline
V8 & Data protection & \changed{16}  &8.1.1 Verify the application protects sensitive data from being cached in server
components such as load balancers and application caches. \\ \hline
V9 & Communication & 8 & 9.1.1 Verify that TLS is used for all client connectivity, and does not fall back to
insecure or unencrypted communications. \\ \hline
V10 & Malicious Code & 10 & 10.2.4 Verify that the application source code and third party libraries do not
contain time bombs by searching for date and time related functions. \\ \hline
V11 & Business Logic & 8 & 11.1.1 Verify that the application will only process business logic flows for the same
user in sequential step order and without skipping steps. \\ \hline
V12 & Files and resources & 15 &12.1.1 Verify that the application will not accept large files that could fill up storage
or cause a denial of service. \\ \hline
V13 & API and Web Service & \changed{13} & 13.1.3 Verify API URLs do not expose sensitive information, such as the API key,
session tokens etc. \\ \hline
V14 & Configuration & \changed{24} &14.1.3 Verify that server configuration is hardened as per the recommendations of
the application server and frameworks in use. \\ \hline

        \end{tabular}
    \end{table*}

\subsection{Motivating Example}
\label{subsec:motivatingExample}

\begin{figure}[!htbp]
	\centering
	\includegraphics[trim = {0 14cm 0cm 0}, clip,width=0.95\textwidth]{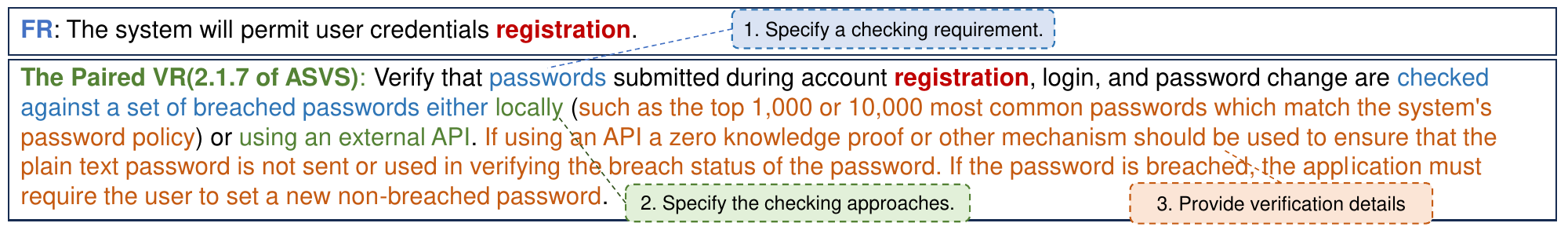}
	\caption{An Example Demonstrating How the Paired VR Specifies Verification Criteria for the Given FR.}
	\label{fig:VRExample}
\end{figure}

To generate specific and actionable SRs methodically, we suggest harnessing relevant VRs from the OWASP ASVS as a roadmap for shaping SRs. This approach is inspired by the insight that ASVS VRs actually outline the security verification benchmarks for FRs in web applications. Take user registration as an instance in Fig. \ref{fig:VRExample}: VR 2.1.7 mandates that engineers should authenticate passwords employing diverse methods. Moreover, it provides explicit instructions for the verification process, such as local checks or the use of external APIs. Armed with this precise guidance, we are confident that LLMs can craft more valid and practical SRs.

Nonetheless, the task of aligning VRs with FRs diverges from conventional synonym matching or question-answering pairing tasks typically encountered in Information Retrieval (IR) and Natural Language Processing (NLP). The complexity arises not necessarily from a direct semantic similarity but often from connections grounded in specific tokens or underlying semantics that may not be immediately apparent. 

As illustrated in Fig. \ref{fig:VRExample}, the FR aligns with the VR because it pertains to user credential registration, signifying the necessity for password validation within the security module—a core objective of the VR. However, this FR and VR pair is neither synonymous nor does it form a classic question-answer duo. Their association is discerned through a single term ``registration'' in the FR and VR.

This simple example prompts us to recognize three critical insights:
\begin{itemize}[leftmargin=0.3cm]
	\item \textbf{Insight I:} The process of retrieving relevant VRs for FRs deviates from traditional information retrieval operations, necessitating domain adaptation.
	\item \textbf{Insight II:} There may be limited semantic overlap between FRs and VRs; often, the linkages are more granular at the token level.
	\item \textbf{Insight III:} Not all tokens in VRs possess equal weight—the term of ``registration'' carry more relevance than ``verify'' in the context of Fig. \ref{fig:VRExample}.
\end{itemize}

The first insight emphasizes the necessity for domain adaptation, a process hindered by the constraints of public FR-VR pairs availability and thereby prompting the synthesis of data. The second and third insights compel us to engineer a custom retriever architecture tailored to our specific needs. The ultimate challenge is to leverage the relevant VRs to steer the generation of high-quality SRs for the specified FRs.

\section{Approach}
\label{sec:approach}

\begin{figure*}[!htbp]
    \centering
    \includegraphics[trim = {1cm 2cm 3cm 0}, clip, width=0.95\textwidth]{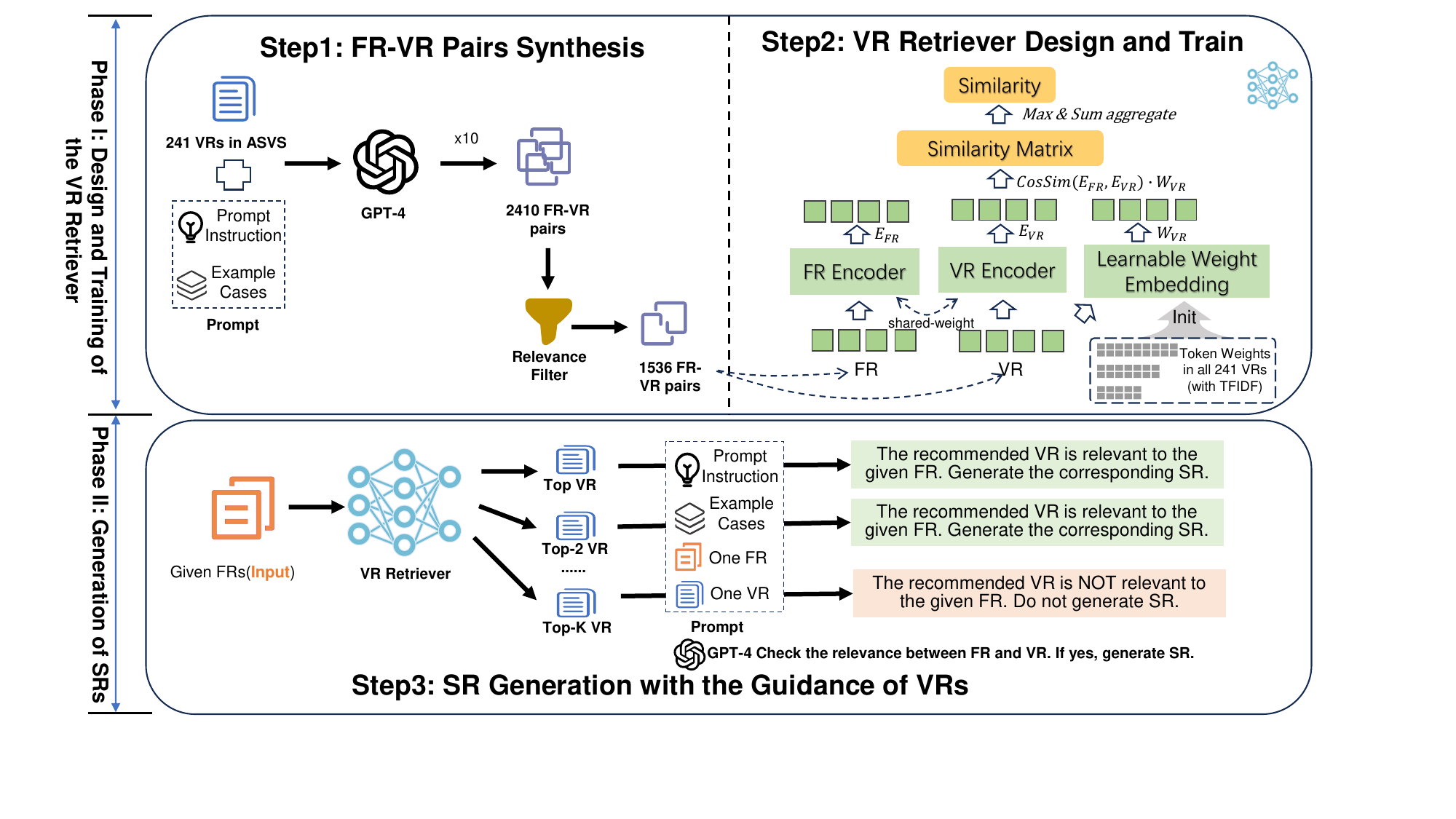}
    \caption{The Overview of \textit{F2SRD}.}
    \label{fig:overview}
\end{figure*}

Fig. \ref{fig:overview} outlines the overview of our approach Functional-to-Security Requirements Derivation (\textit{F2SRD}). It includes two phases. 

Phase I entails the development and training of a VR retriever, which is designed to pinpoint VRs pertinent to specified FRs. This phase encompasses two fundamental steps. Initially, due to the scarcity of FR-VR pairings, we utilize GPT-4 to synthesize 10 unique FRs for each of the 241 VRs listed in the ASVS. Following a relevance filtering process, this yields a total of 1536 viable synthesized pairs. Subsequently, we construct a specialized retrieval architecture featuring a learnable weight embedding layer that assigns variable importance to tokens within the VRs during the similarity assessment. The retriever is then trained and validated on the dataset synthesized in the first step.

In Phase II, the \textit{F2SRD} model is deployed in a practical setting, tasked with strengthening security by refining actual FRs into detailed functional security requirements. Utilizing the trained retriever from Phase I, it identifies the top-k relevant VRs corresponding to each FR. This identification suggests potential weaknesses within the FR that should be addressed. The utilized LLM, specifically GPT-4 in this study, assesses the need for security verification of the FRs—ascertaining whether they are security-critical in relation to the identified VRs. Should verification be warranted, the LLM then generates the pertinent SRs.

\begin{figure}[htbp]
    \includegraphics[trim = {0 3.5cm 6.3cm 0}, clip, width=0.95\textwidth]{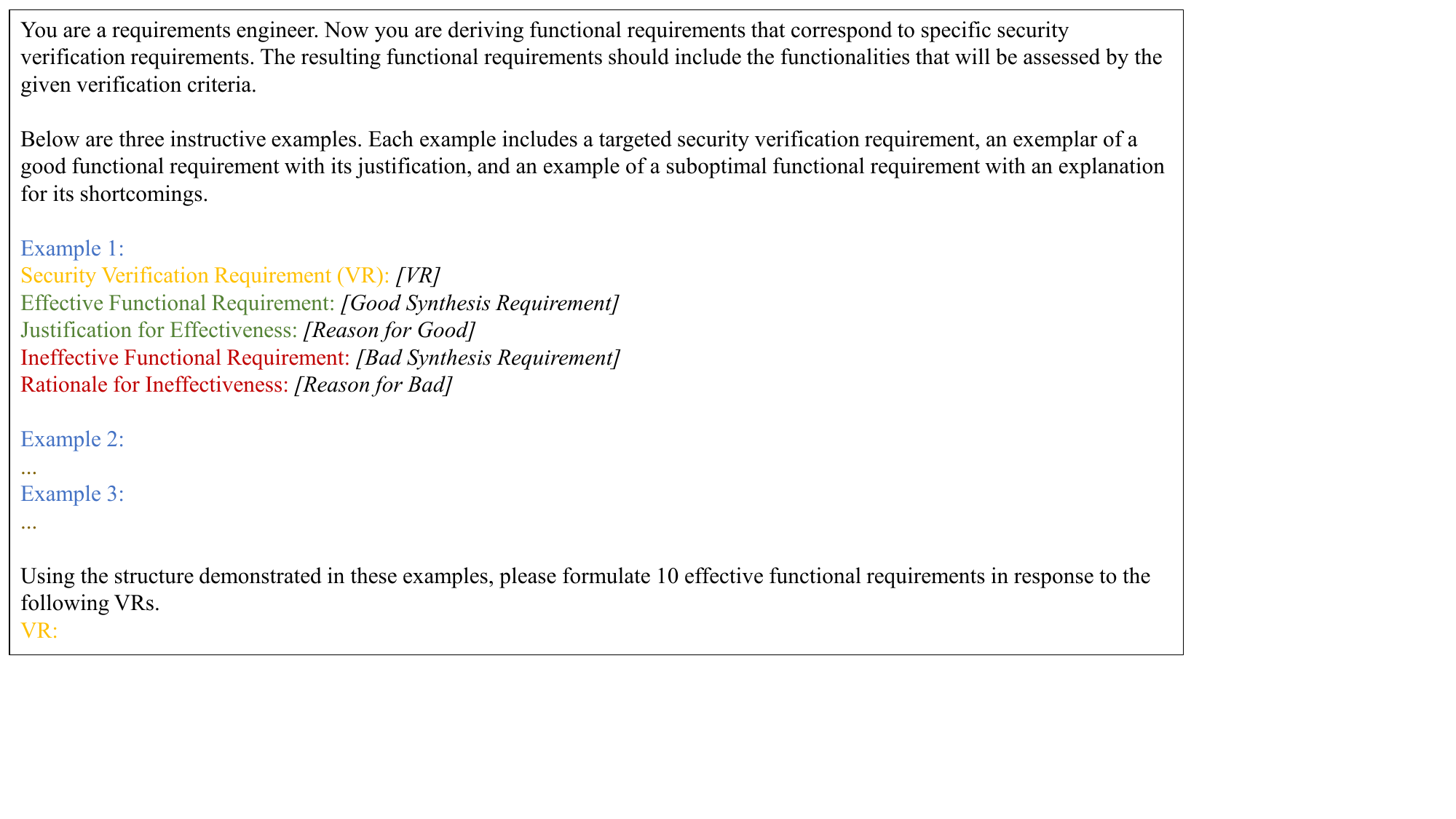}
    \caption{\changed{Data Synthesis Prompt Template Used in Step 1 of F2SRD.}}
    \label{fig:synthesis_prompt}
\end{figure}

Phase I includes two steps. 
\noindent \textbf{Step 1: FR-VR Pairs Synthesis.} 
Drawing inspiration from Jeronymo et al. \cite{jeronymo2023inpars}, we adopt prompt engineering with LLMs to generate synthetic training data. Our prompt template for crafting FR-VR pairs (as shown in Fig. \ref{fig:synthesis_prompt}) comprises an instruction set along with three exemplifying cases. Following this template, GPT-4 is tasked with generating FRs that require verification in light of a specified VR, specifically because these FRs encompass functions addressed by the VR.

In contrast to the approach of Jeronymo et al., our prompt design encompasses examples of both well-conceived and poorly-conceived synthesis requirements, complete with explanations delineating their strengths and weaknesses, respectively. This structure enables GPT-4 to more accurately discern the standards for successful synthesis. A similar concept has been substantiated in code repair applications \cite{zhao2024novelapproachautomateddesign, zhao2024enhancingllmbasedautomatedprogram}. The ``Reason for good'' section demonstrates that the sampled FR presents vulnerabilities that are well-described by the VR. Conversely, the ``Reason for bad'' details three categories of inadequate synthesis identified in our empirical analysis. These pitfalls include: 1) synthesized data that fail to serve as FRs, i.e, describing the system or system element functions or tasks to be performed by the system \cite{8463987}; 2) synthesized content that is too abstract and lacks specific detail; 3) synthesized examples that explicitly reveal VR-related vulnerabilities rather than integrating them with subtlety. To curb such unsatisfactory outcomes, each type of flaw is demonstrated through a dedicated negative example included in the prompt.
We contend that the incorporation of such detailed explanations will significantly elevate the fidelity of the generated FRs.

We filter out 36 VRs from the ASVS that pertain to categories like Secure Software Development Lifecycle (Section 1.1 of ASVS) and Build and Deploy (Section 14.1 of ASVS) as they focus on the development process rather than software functionality, leaving 241 VRs (from the original 277 valid VRs). \changed{To preserve all relevant details, each VR is constructed by concatenating its chapter title, section title, and description using hyphens. The resulting VRs range from 8 to 103 tokens, with an average of 33 tokens.  We then synthesize 10 FRs for each VR, generating a total of 2,410 FR-VR pairs.}

To further refine the quality of generated FR-VR pairs, we adopt a zero-shot retriever—specifically ColBERTv2 \cite{santhanam2021colbertv2}—as a quality control mechanism, in a manner akin to Saad-Falcon et al.\cite{saad-falcon-etal-2023-udapdr}. We select ColBERTv2 for this task due to its enhanced capabilities in information retrieval compared to LLMs, which more precisely aligns with our Phase I training objective: to cultivate a retriever skilled in pinpointing VRs pertinent to any specified FR. A more accurately matched dataset translates to improved training performance.

\changed{We exclusively utilize data that meets our stringent criteria. For each generated FR-VR pair, we employ zero-shot ColBERTv2 to query VRs based on the FR from a pool of 241 VRs. A synthetic FR-VR pair is deemed acceptable only if this VR ranks within the top 30\% of ColBERTv2’s retrieval results for the given FR. We apply this filtering strategy because it has been validated by existing methods \cite{saad-falcon-etal-2023-udapdr, jeronymo2023inpars, dai2022promptagator, bonifacio2022inpars}}. Upon enforcing these criteria, we have curated a dataset consisting of 1,536 pairs of high quality.

\noindent \textbf{Step 2: FR to VR Retriever Training.} In this step, we design and train a tailored retriever. We opt for ColBERTv2 as the underlying backbone due to its advanced performance as a retriever in IR \cite{santhanam2021colbertv2}. Besides, its late interaction strategy matches our scenario. This approach represents texts as multi-dimensional token-level vectors instead of a singular vector representation, utilizing MaxSim for aggregation to compute similarity scores. This method is particularly adept at facilitating matches based on specific tokens or semantics, which resonates with the second insight from Section \ref{subsec:motivatingExample}.

The first insight from Section \ref{subsec:motivatingExample} drives us to embed the pre-existing knowledge of ASVS into our retriever. Yet, the entirety of the 241 VRs within ASVS surpasses the input size limit of ColBERTv2, presenting a challenge in modeling the full spectrum of ASVS.
To circumvent this, we pre-calculate token weights from ASVS and embed them into our retriever. Concretely, we employ the TF-IDF model to determine and normalize these weights. Subsequently, we introduce a learnable weight embedding layer into the architecture of ColBERTv2, which is initialized with these token weights calculated with TF-IDF. The parameters of this layer are subject to updates during the training process to facilitate further adaptation. The comprehensive architecture of the retriever is delineated in Step 2 of Fig. \ref{fig:overview}.

The retriever computes similarity score as follows: given a functional requirement $FR$ and a verification requirement $VR$ with lengths $m$ and $n$ respectively, we first obtain $E_\text{FR}$ and $E_\text{VR}$ using shared-weight encoders, representing their token-level embeddings. Second, we use the previously mentioned weight embedding layer to obtain token weights in the VR, referred to as $W_\text{VR}$. Third, we calculate the token-level cosine similarity between the FR and VR and weight it using $W_\text{VR}$ to get the similarity matrix, denoted as $Sim_\text{matrix}$. Finally, we aggregate the similarity score using a method similar to ColBERTV2. The specific formulas are as follows:
\begin{align}
    &E_{FR}, E_{VR} = Encoder_{FR}(FR), Encoder_{VR}(VR) \\
    &W_{VR} =  Embedding_{weight}(VR) \\
    &Sim_{matrix} = W_{VR} \cdot CosSim(E_{FR}, E_{VR}) \\
    &Sim_{FR, VR} =  \sum_{i\in m}(\max\limits_{j \in n}Sim_{matrix}) 
\end{align}
We utilize the 1,536 FR-VR pairs derived from Step 1 as our dataset, adopting a 9:1 ratio for training and validation. The effectiveness of this retrieval approach is assessed in Section \ref{sec:RetriverEffectiveness}, and it is subsequently utilized in Step 3.

\noindent \textbf{Step 3: Generate SRs for FRs.} Step 3, the only step in Phase II, represents the practical application stage of \textit{F2SRD}. In this phase, \textit{F2SRD} takes FRs as inputs and outputs the corresponding SRs. The process begins with \textit{F2SRD} deploying the trained retriever to identify the top-K VRs that are most relevant to each FR. Subsequent to retrieval, each FR-VR pair is incorporated into a SR generation prompt and fed into GPT-4.

Two scenarios are considered at this juncture. In the first, if an FR truly requires verification by the identified VR, GPT-4 should craft an SR informed by the VR and FR; conversely, in the second scenario, if an FR is unrelated to the retrieved VR (indicating an inapplicable VR for the given FR), GPT-4 refrains from generating an SR. This selective approach is intended to minimize unnecessary computational overhead and reduce the burden on human reviewers. \changed{It is important to note that in certain special cases, an FR can also serve as an SR. However, this does not imply that no additional SRs are necessary. For example, a requirement related to login and password verification involves associated SRs such as password storage and logout management. This scenario still adheres to the principle of our approach, which is to determine whether additional VRs and SRs need to be implemented.}

\begin{figure}[htbp]
    \includegraphics[trim = {0 3cm 3cm 0}, clip, width=0.95\textwidth]{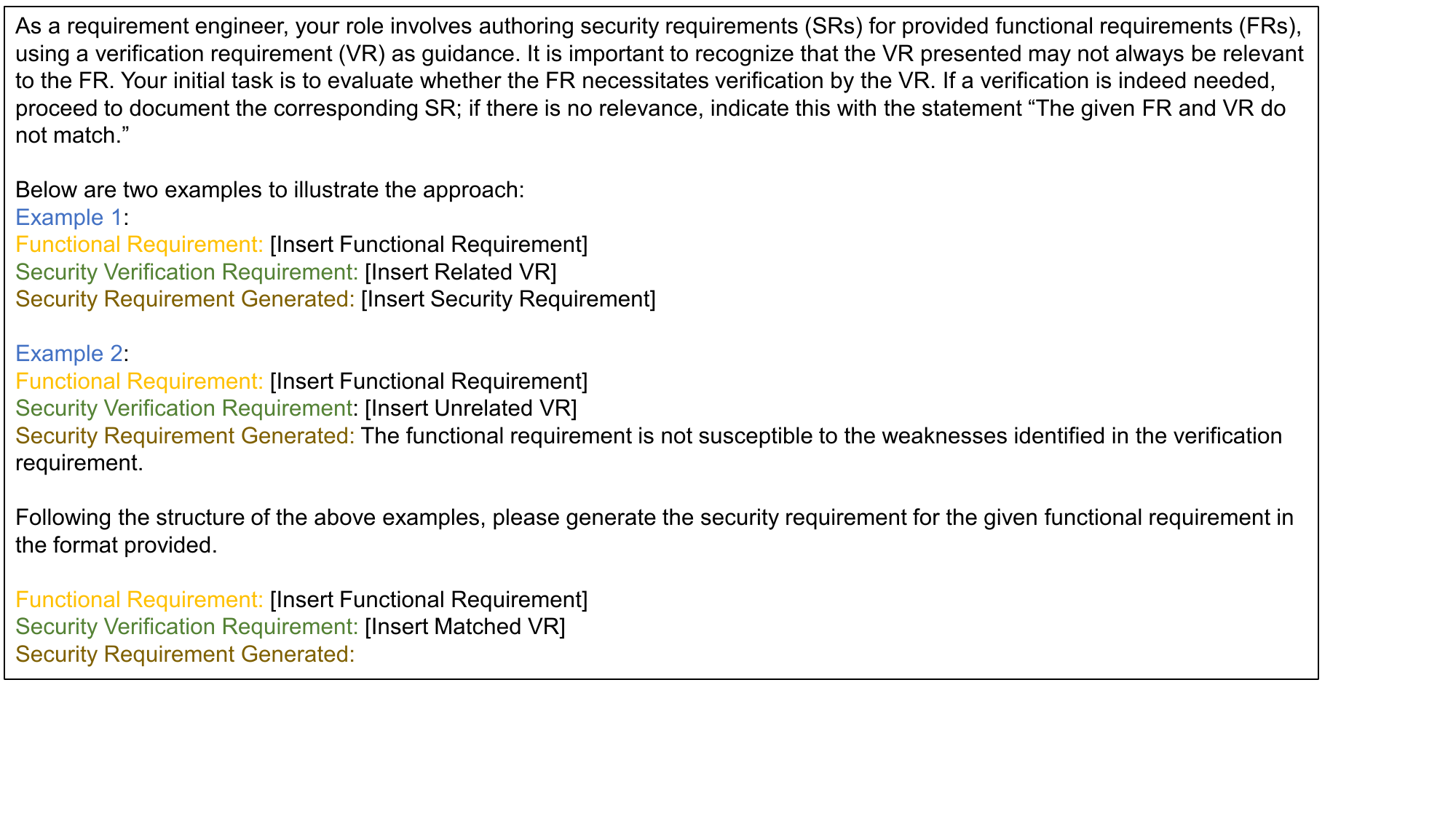}
    \caption{\changed{SR Generation Prompt Used in Step 3 of F2SRD.}}
    \label{fig:generation_prompt}
\end{figure}

\changed{The template for generating SRs is illustrated in Fig. \ref{fig:generation_prompt}. This template includes task-specific instructions and two manually composed examples. These examples demonstrate two scenarios: one where the recommended VR is relevant to the given FR, and another where it is not. Consequently, the corresponding SR should be generated only in the first example.}

\section{Evaluation}
\label{sec:eva}

We assess the efficacy of our \textit{F2SRD} by addressing the following two research questions:

\noindent \textbf{RQ1 (Effectiveness of VR Retriever)}: How effectively can our VR retriever recommend relevant and appropriate VRs for specified FRs?

\noindent \textbf{RQ2 (Effectiveness of F2SRD):} How does the performance of our \textit{F2SRD} in generating inspired, diverse, and specific SRs compare against two established baselines from the given FRs?

\subsection{Datasets}
\label{sec:datasets}
We use datasets provided by the 25th IEEE International Requirements Engineering Conference (RE ’17) call for data track papers\footnote{\url{http://re2017.org/pages/submission/data papers/}}, which are widely used \cite{fong2018software, houmb2010eliciting}. These datasets exhibit characteristics of web application and fall within the scope of ASVS applicability. \changed{We utilize all FRs from these datasets directly. These FRs vary in length from 9 to 127 tokens, with an average of 33 tokens.}

\noindent\textbf{ePurse}: The ePurse (Common Electronic Purse) dataset contains 41 FRs, mainly describing the features of an electronic wallet application. These requirements specify various functions and characteristics of electronic wallets, such as account management and payment transactions.

\noindent\textbf{CPN}: The CPN (Customer Premises Network) dataset contains 169 FRs related to various needs and specifications of customer premises networks. This dataset provides detailed descriptions of network topology, device configuration, and quality of service requirements.

\noindent\textbf{GPS}: The GPS (Global Platform Specification) dataset comprises 113 FRs related to the interoperability, security, and management aspects of smart cards and their associated applications.

\subsection{Addressing RQ1: Effectiveness of VR Retriever}
\label{sec:RetriverEffectiveness}

The foundational effectiveness of the VR retriever is crucial for generating accurate and relevant prompts for SR generation, which in turn determines the quality of the resultant SRs. Consequently, we prioritize the evaluation of the VR retriever's effectiveness.

\changed{The objectives of this evaluation are twofold: \ding{192} to gauge the reasonableness of these pairings, specifically determining the necessity of the recommended VRs for verifying their corresponding FRs; and \ding{193} to evaluate the effectiveness of the two key steps of our retriever—training data filtering (Step 1) and retriever training (Step 2).}

\changed{To achieve these evaluation objectives, we implemented two additional variants of our retriever by removing the training data filtering in one and the training itself in the other. These two variants, along with our original retriever, were tasked with recommending five VRs for each FR across three distinct projects, resulting in a total of 1,615 FR-VR pairings from 323 FRs (i.e., all FRs in the three datasets). }


\changed{Due to the lack of ground truth for VR recommendations, manual evaluations were necessary. Adhering to minimal sampling standards, we randomly selected 214 FR-VR pairs from the results of each retriever, ensuring a 95\% confidence interval for our study. To avoid redundant and unnecessary annotations due to potential overlap among the results of different retrievers, we constructed their union set. This process resulted in a total of 470 unique FR-VR pairs.}
These samples were then divided into eight groups and reviewed by \changed{fourteen} participants from a master's-level software engineering (SE) class at a prestigious university. Each participant has over four years of SE experience, fundamental knowledge of software security, and an understanding of the importance of security requirements.



\changed{The task of the annotators was to assess the relevance or irrelevance of each FR-VR pair. In this context, relevance means that the FR necessitates this VR for validation. In other words, relevance indicates that this VR exposes potential vulnerabilities or issues within the FR.
To ensure the annotators fully understood our task and the concept of relevance, we provided a tutorial. In this tutorial, we introduced our annotation task and explained two positive examples (relevant pairs) and one negative example (irrelevant pair). Additionally, we discussed the reasons for relevance and irrelevance with the participants to further clarify the criteria.} It was explicitly communicated that our aim was to acquire `genuine' annotations regarding the relevance between the FR and VR in each pair. Additionally, to maintain impartiality, the annotators were not affiliated with our research team, thereby eliminating any bias towards providing `relevance' annotations for our dataset. Following their independent annotations, the two participants reviewed the same data set and engaged in a joint discussion to resolve any discrepancies until consensus was reached.

We used the finalized annotations from this collaborative review as the gold standard for our retriever's  evaluation. We calculated the retriever's accuracy by determining the proportion of `relevant' samples it identified compared to the gold standard. The results are presented in Table \ref{tab:RetriverAccuracy}.
\changed{We can make the following two observations.}
\begin{itemize}[leftmargin=0.4cm]

    \item \changed{\textbf{Our retriever is effective, with the average accuracy of 80.4\%.} With the datasets of ePurse and CPN, the retrieval accuracy is above 84\%. Meanwhile, we can also observe that the accuracy on the GPS dataset was lower, with 70.3\%. This may be attributed to the higher incidence of cross-references among FRs within the GPS dataset. For example, one FR in GPS states: ``\emph{The concept of the Life Cycle of the card (see section 5.1 - Card Life Cycle) may be used to determine the security level of the communication between the card and an off-card entity.}'' Without a complete understanding of the \emph{card's life cycle}, accurately identifying security vulnerabilities and matching the appropriate VRs for this FR become challenging. Indeed, without a complete understanding, formulating accurate recommendations is undoubtedly problematic.} 
    
    \item \changed{\textbf{Both the training and data filtering designs are indispensable, with the influence of training being larger than that of data filtering.} Specifically, when these steps were removed, our retriever achieved average accuracy gains of 34.45\% and 8.94\%, respectively. Even without data filtering (as shown in the row "w/o data filtering"), training the retriever directly with synthesized data still improves performance compared to the zero-shot ColBERTv2 (as shown in the row "w/o training"). This underscores the critical importance of domain-specific training. This aligns with Insight 1 (discussed in Section \ref{subsec:motivatingExample}), which suggests that a general-purpose retriever struggles to align FRs with relevant VRs, resulting in a substantial drop in overall performance. Additionally, data filtering on the training dataset, even when using an existing retriever like ColBERTv2, can enhance the quality of the final retriever. This observation can inspire related research involving retrievers, potentially extending beyond our FR-VR retrieval context.}
\end{itemize}

\begin{table}[htbp]
\caption{\changed{Accuracy of the retriever on three datasets.}}
\centering
\begin{tabular}{|l|l|l|l|l|} 
\hline
\multirow{2}{*}{Approach}                        & \multicolumn{3}{c|}{Accuracy}                  & \multirow{2}{*}{Avg. Accuracy}  \\ 
\cline{2-4}
                                                 & ePurse        & CPN            & GPS           &                                 \\ 
\hline
Retriever in F2SRD                                            & \textbf{84.6\%}(22/26) & \textbf{86\%}(98/114)   & \textbf{70.3\%}(52/74) & \textbf{80.4\%}(172/214)                 \\ 
\hline
w/o training                                     & 69.2\%(18/26) & 64\%(73/114)   & 50\%(37/74)   & 59.8\%(128/214)                 \\ 
\hline
w/o data filtering & 80.8\%(21/26) & 78.1\%(89/114) & 64.9\%(48/74) & 73.8\%(158/214)                 \\
\hline
\end{tabular}
\label{tab:RetriverAccuracy}
\end{table}


To assess the level of agreement among participants' annotations, we utilized the Intraclass Correlation Coefficient (ICC), as outlined by \cite{Fisher1992}. The ICC is a widely recognized statistical tool for gauging interrater reliability. In our study, we implemented a two-way random effects model assessing consistency across multiple raters, denoted as ICC(2,k). Contrary to our expectations, the ICC value obtained was only \changed{0.47}, with a highly significant p-value of \changed{$10^{-14}$}. This reflects merely \emph{moderate consistency} \cite{ICCGuideline}, as established by the standard 95\% confidence interval.

This outcome resonates with the inherent complexity faced by requirement\changed{s} engineers when trying to discern the relevance between FRs and VRs. On one hand, interpreting VRs necessitates a solid grasp of cybersecurity principles, which poses a challenge for requirements engineers. More importantly, even if a VR could potentially enhance the security aspects of an FR (hence being relevant), the linkage between the two might not always be immediately evident.

\begin{figure}[!htbp]
	\centering
	\includegraphics[trim = {0 14.7cm 4cm 0}, clip,width=0.95\textwidth]{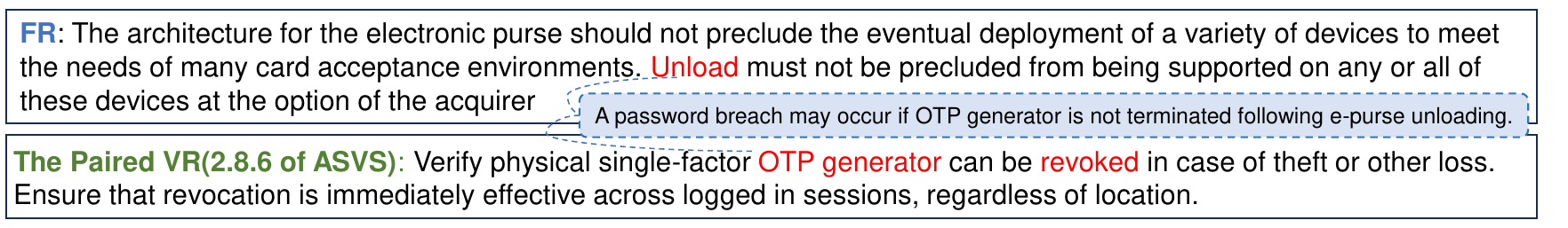}
	\caption{An Example of Indirect Yet Important Relevance Between  Between FR and VR.}
	\label{fig:RetrieverExample}
\end{figure}

Take, for instance, the scenario depicted in Fig. \ref{fig:RetrieverExample}, where the FR indicates support for uninstallation within the e-Purse system, seemingly unrelated to the ``OTP generator'' mentioned in the VR. However, a security risk arises if the OTP (One-Time Password) generator remains after uninstallation, or if session authentication persists, potentially resulting in a serious password breach. \changed{To provide a clearer illustration, we have applied manually annotated color-coding to highlight key information in the figures.} Accurate annotation in such cases demands that participants meticulously consider various subtleties, backed by domain expertise pertaining to both the FR and the recommended VRs, underscoring the task's difficulty.

Additionally, we have assessed the accuracy of human annotations by comparing them to their consensus answers following discussion, with the average accuracy reaching \changed{85.1\%}. While this figure surpasses that of our retrieval system, the margin is not particularly striking. Furthermore, given that engineers must evaluate each FR against over 200 VRs, the task becomes exceedingly labor-intensive. We are confident that our retrieval system can alleviate the burden of this extensive analysis for engineers while maintaining an accuracy level on par with that of human counterparts.

\subsection{Addressing RQ2: Effectiveness of F2SRD}
\label{sec:effectiveEva}


\subsubsection{Baselines} 
\label{sec:baselines}
We selected two state-of-the-art approaches as our baselines.

\noindent\textbf{RelGAN} (Relational Generative Adversarial Network): This is a novel generative adversarial network (GAN) architecture for text generation \cite{nie2018relgan}. Koscinski V et al.\cite{koscinski2023demand} utilized it to synthesize SR specifications, and their research demonstrated RelGAN's capability in generating realistic and practical SRs. Therefore, RelGAN can serve as a baseline model.

\noindent\textbf{ChatGPT} (Chat Generative Pre-trained Transformer): Released by OpenAI, ChatGPT is one of the state-of-the-art models with a strong advantage in generative tasks. 
It has established itself as a formidable baseline across a myriad of generative applications \cite{zhao2024novelapproachautomateddesign}. In our work, we adopt GPT-4 as the foundational benchmark.

\subsubsection{Implementation Details}
\label{sec:implementation}
\noindent\textbf{F2SRD}: Building upon the model trained as per \cite{santhanam2021colbertv2}, we retain the original ColBERTv2's default learning rate of 3e-5. For the newly introduced Learnable Weight Embedding Layer, we set the learning rate to 6e-7, which is fine-tuned during the training process using the FR-VR pairs.
We recall the top-5 VRs for each FR and attempt to generate one SR for each FR-VR pair using the GPT-4 model.

\noindent\textbf{RelGAN} \cite{nie2018relgan}: RelGAN is a non-directional generative model that does not need FRs as input during SRs generation. We incorporate the source code from Nie et al. \cite{nie2018relgan}, which is also utilized by Koscinski et al. \cite{koscinski2023demand}. All parameters specified in \cite{koscinski2023demand} are retained in our replication.
In alignment with Koscinski et al. \cite{koscinski2023demand}, it becomes necessary to train RelGAN for each individual project. The training dataset incorporates FRs from the project currently under test, as well as both FRs and SRs from two additional datasets in our selected ones. Subsequently, SRs are generated at a volume quintuple to that of the FRs within each dataset.

\noindent\textbf{GPT-4}: We employ the dialogue interface of GPT-4 for the generation of SRs. In accordance with the configuration established by \textit{F2SRD}, we generate five SRs corresponding to each FR. The specific prompt utilized is attached in Fig. 10 of the replication package.

\subsubsection{Evaluation Aspects}
\label{sec:metrics}

The SRs generated are expected to exhibit three key characteristics:

\noindent \textbf{Inspiration}: SRs ought to serve as a catalyst for requirements engineers, prompting the derivation of innovative and critical SRs based on the given FRs.

\noindent \textbf{Diversity}: To ensure comprehensive security, it is critical that generated SRs address a broad array of protections. Aligning with standards such as OWASP ASVS \cite{ASVS} and NIST guidelines\footnote{\url{https://nvlpubs.nist.gov/nistpubs/FIPS/NIST.FIPS.200.pdf}}, not only fortifies against diverse threats but also reinforces stakeholder trust in the system's integrity.

\noindent \textbf{Specificity}: Each SR should be articulated with precision, eschewing any ambiguity. This characteristic complies with the requirements of ISO 29148\cite{iso2011ieee} and is supported by numerous studies \cite{bulusu2017towards}\cite{lamsweerde2009requirements}.

We perform quantitative assessment to measure the levels of \emph{inspiration} and \emph{diversity}, while human evaluation is utilized specifically for analyzing \emph{inspiration} and \emph{specificity}. Given its critical importance to our task, \emph{inspiration} is evaluated through dual methodologies, as it encompasses elements of both ``unexpectedness'' and ``reasonableness.'' The component of unexpectedness is quantified using the metric of \emph{self-information}, whereas reasonableness is appraised through manual evaluation processes.

\subsubsection{Quantitative Evaluation: Metrics}

\begin{itemize}[leftmargin=0.3cm]
    \item \textbf{Self-information (SI)}: We employ self-information \cite{katona1976huffman} to measure the \emph{unexpectedness}  of the generated SRs (one dimension of \underline{inspiration}). An established concept from information theory \cite{informationTheory}, it evaluates the level of surprise or uncertainty associated with the issuance of a specific SR. Under Shannon's theoretical framework \cite{6773024}, an SR's rarity directly contributes to its self-information, with rarer instances indicating higher unexpectedness. The formula is as follows:

    \begin{equation}
    SI(SR)=
    \begin{cases}
    -\log_2{p(SR)}& \text{ $ SR \in Project_{input} $ } \\
    0& \text{ $ SR \notin Project_{input} $ }
    \end{cases}
    \end{equation}

    We calculate the SI of each SR using an autoregressive generation model, opting for the Davinci-002 architecture \cite{ye2023comprehensive} due to its remarkable performance in generating coherent and contextually relevant text. 
    It should be noted that some experimental approaches, particularly RelGAN, may yield SRs that fall outside the intended project scope. For such instances, we assign an SI value of 0 to these out-of-scope SRs. We calculate the average value as the final result of one approach towards one specific project.
    
    \item \textbf{Self-BLEU}: To gauge the \underline{diversity} of SRs generated for each unique project, we utilize the Self-BLEU metric \cite{zhu2018texygen}, which is tailored to measure the variance among generated datasets. This metric is particularly applied to assess the range of SRs produced for \emph{distinct} FRs, reflecting the unique security context required by each FR. A lower Self-BLEU score signifies a larger discrepancy among SRs, denoting a higher level of diversity. The formula is defined as follows: 
    \begin{equation}
    SelfBLEU_{i} = 
    \begin{cases}
    BLEU(SR_{i},SR_{1\ldots i-1,i+1 \ldots N})& \text{ $ SR \in Project_{input} $ } \\
    1& \text{ $ SR \notin Project_{input} $ }
    \end{cases}
    \end{equation}

    $SR_{i}$ represents one SR generated from the $i$th FR for the input project. To assess the diversity of SRs across various FRs, and considering that multiple SRs derived from the same FR typically exhibit variations, we randomly select one representative SR from each set for evaluation.
    $N$ represents the total number of FRs. Since the SR out of the project scope does not make any sence, we assign the BLEU value as the maximize one, i.e., 1. The final result for each approach on a specific project is determined by taking the average value, similar to SI.

    \item \textbf{Vocabulary size (VS)}: In addition to assessing the overall diversity of SR descriptions, we measure lexical diversity using \emph{vocabulary size (VS)}, which reflects the richness of the vocabulary incorporated within the generated SRs. We compute VS by tallying the unique words produced by a specific approach in a particular project's SRs.
    For the SRs that fall outside the scope of the project, we omit their vocabulary from the VS calculation.
\end{itemize}

\subsubsection{Post-filtering for RelGAN}
Owing to the inherent technical principles of RelGAN, it is prone to generating SRs that may be relevant to other projects included in the training dataset. For example, it could produce an irrelevant SR like ``offline authentication must be conducted between the PSAM and card at any point during transactions by the load acquirer'' for a GPS project, which has no connection to \emph{GPS} technology but clearly pertains to the \emph{ePurse} project. Requirements of this nature contribute nothing beyond confusion to the task of SRs specification for engineers. Consequently, our first step prior to evaluation is to sift out these types of requirements.

We implement a meticulous automated filtering approach that targets unique proper nouns associated with specific projects. This process is further refined through detailed human review. In our filtering procedure, we zero in on the project names and proper nouns—especially those distinguished by initial capital letters—as the targeted keywords for identification. Essentially, the presence of these select terms signals an association with particular projects.
The diversity of domains represented by the selected projects enhances the precision and effectiveness of this keyword-centric filter.

Table \ref{tab:RelGANFiltering} provides details on indicative keywords, the number of SRs generated, and the degree of filtering applied. On average, approximately 38.27\% of SRs produced by RelGAN are subject to filtration. Notably, for projects such as \emph{ePurse} and \emph{GPS}, fewer than half of the generated SRs remain after filtering, with only 47.3\% and 33.1\% respectively. Given this high rate of exclusion, the efficacy of RelGAN in synthesizing SRs may be called into question.

\begin{table}[htbp]
\small
\caption{Filtered SR Outcomes from RelGAN Generation.}
\begin{center}
\renewcommand{\arraystretch}{1.2}
\begin{tabular}{|c|c|c|c|c|}
\hline
\textbf{Dataset} & \textbf{Keywords} & \textbf{\#All} & \textbf{\#Removed} &\textbf{\#Left} \\
\hline
ePurse & ePurse, pos, psam, ceps, cep & 205&108&97 \textcolor{blue}{(47.3\%)} \\ \hline
CPN & cpn, cng &845&132&713 \textcolor{blue}{(84.4\%)}\\ \hline
GPS &gps, globalplatform&565&378&187 \textcolor{blue}{(33.1\%)}\\
\hline
\end{tabular}
\label{tab:RelGANFiltering}
\end{center}
\end{table}

\subsubsection{Quantitative Results}
\label{subsubsec:quantitativeResults}

The results of our quantitative evaluation, focusing on inspiration (as measured by self-information) and diversity (as gauged through self-BLEU scores and vocabulary size), are presented in Table \ref{tab:quantitativeResults}. To facilitate easier comparison, we have emboldened the top-performing values for each metric within their respective projects. It is evident that our proposed \textit{F2SRD} outperforms all competitors in each of the three metrics across all three projects. To more clearly demonstrate the effectiveness of \textit{F2SRD}, we performed Welch's t-test \cite{1947The} on the metrics of self-information and self-BLEU. Vocabulary size is not factored into the analysis because of the limited number of samples. Each approach results in a single vocabulary size value per project.

The results (Table \ref{tab:quantitativeTtest}) show that our method significantly outperforms the two baselines with a 95\% confidence level. Among the baseline models assessed, GPT-4 excels in eight of the nine evaluated categories, demonstrating its strengths than RelGAN. This is also the reason that we select GPT-4 as our backbone model.

The higher Self-Information exception in CPN, as observed with GPT-4, can be attributed to the extensive scale of training samples utilized for RelGAN fine-tuning, coupled with a substantial reserving ratio (84.4\%, as shown in Table \ref{tab:RelGANFiltering}). Specifically, an increased number of training samples within CPN prompts the model to generate text that more closely aligns with the project's requirements, ultimately leading to the preservation of a greater proportion of these requirements. Moreover, the generated text encompasses specialized terms and combinations thereof which are pertinent to this particular project. The incorporation of these terms occurs with a comparatively lower frequency than more common counterparts, resulting in a higher degree of expectedness. Nonetheless, for other evaluated metrics, its performance is less impressive.


\begin{table}[htbp]
\small
\caption{Quantitative Evaluation Results for SR Generation.}
\begin{center}
\renewcommand{\arraystretch}{1.2}
\begin{tabular}{|c|c|c|c|c|}
\hline
\multirow{2}{*}{\textbf{Dataset}} & \multirow{2}{*}{\textbf{Approaches}} & \textbf{Inspiration} &\multicolumn{2}{c|}{\textbf{Diversity}} \\ \cline{3-5}
&&\textbf{Self-Information} & \textbf{Self-BLEU} &\textbf{Vocabulary size} \\ \hline
\multirow{3}{*}{ePurse} & RelGAN &90.03 &0.75 &440\\ 
&GPT-4 &144.98 & 0.66 &549\\ 
&F2SRD& \textbf{169.09} & \textbf{0.53} &\textbf{737}\\ \hline

\multirow{3}{*}{CPN} &RelGAN &160.70 &0.87 &800\\ 
&GPT-4&141.17 &0.78 &885\\ 
&F2SRD &\textbf{182.48}& \textbf{0.73} &\textbf{1014}\\ \hline

\multirow{3}{*}{GPS} & RelGAN &71.58 &0.91 &527\\ 
&GPT-4 &161.69 & 0.78 &773 \\
&F2SRD &\textbf{227.05} & \textbf{0.72} &\textbf{1202} \\ \hline
\end{tabular}
\label{tab:quantitativeResults}
\end{center}
\end{table}

\begin{table}[htbp]
\small
\caption{Welch's t-test Results of Quantitative Results (p-value).}
\begin{center}
\renewcommand{\arraystretch}{1.2}
\begin{tabular}{|c|c|c|c|}
\hline
\textbf{Dataset} & \textbf{Comparison Groups} & \textbf{Self-Information} &{\textbf{Self-BLEU}} \\ 
\hline
\multirow{2}{*}{ePurse} & F2SRD VS. RelGAN &$10^{-17}$ &$10^{-5}$ \\ 
&F2SRD VS. GPT-4 & $10^{-7}$& $10^{-9}$\\  \hline

\multirow{2}{*}{CPN} & F2SRD VS. RelGAN &$10^{-8}$ &$10^{-9}$ \\ 
&F2SRD VS. GPT-4 & $10^{-82}$& $10^{-3}$\\  \hline

\multirow{2}{*}{GPS} & F2SRD VS. RelGAN &$10^{-102}$ &$10^{-21}$ \\ 
&F2SRD VS. GPT-4 & $10^{-57}$& $10^{-9}$\\  \hline
\end{tabular}
\label{tab:quantitativeTtest}
\end{center}
\end{table}

\subsubsection{Human-Subject Study: Study Design}
\label{subsec:humanStudy}

To evaluate whether the generated SRs effectively serve requirements engineers in swiftly crafting specific SRs, we conducted a human-subject study to assess their inspiration and specificity aspects. Given the frequent shallow issues with RelGAN-generated outputs—such as syntactic problems present in 21\% of cases \cite{koscinski2023demand} and out-of-scope content occurring in 38.27\% of instances (as seen in Table \ref{tab:RelGANFiltering})—our human evaluation focused solely on samples produced by GPT-4 and our \textit{F2SRD} model.

As outlined in Section \ref{sec:datasets}, the three datasets combined comprise a total of 323 FRs. Our \textit{F2SRD} framework has successfully generated 1550 SRs, having filtered out 115 FR-VR pairs deemed unreasonable. Employing the minimum sampling size theorem, we determined a sample size of 213 SRs from our output is required for a 95\% confidence level with a margin of error of 5\%. We randomly selected this sample from our pool of generated SRs. This approach allowed us to define the range of target FRs and the corresponding SRs for each FR, given the maintained traceability between them. For each target FR, we also randomly selected an equivalent number of SRs generated by GPT-4 to match our sample size, resulting in 213 tuples. Each tuple consists of one FR along with one corresponding SR from both GPT-4 and \textit{F2SRD}. These tuples were divided into 11 sets, with the first 10 containing 20 tuples each, and the final set comprising 13. We anonymized the source of the SRs within each tuple and randomized their order to prevent any potential bias.

Our questionnaire design incorporates Likert-scale dimensions for both Specificity and Inspiration. Specifically, we utilized a five-point scale for specificity (1-very unspecific, 2-somewhat unspecific, 3-neutral, 4-somewhat specific, 5-very specific), complemented by a three-point scale for inspiration (1-participant could easily think of similar SRs upon reading the FR, 2-analyst would need to give some thought before generating similar SRs, 3-analyst would find it difficult to come up with similar SRs). It is important to note that we used only positive scales for inspiration due to its lack of negative attributes.

We invited 22 participants, all with a minimum of four years of experience in software engineering either academically or professionally. To maintain consistency, two participants evaluated the same set of tuples from one bucket independently using the questionnaires designed.

\subsubsection{Results}
\label{sec:effectiveResults}

To assess the agreement between two independent sets of results for the 213 tuples, we employ the ICC \cite{Fisher1992} for gauging interrater reliability. As in Section \ref{sec:RetriverEffectiveness}, we apply ICC(2,k). The resultant ICC value is 0.82, accompanied by a p-value of $10^{-181}$, which is substantially lower than the standard threshold of 0.05. This denotes \emph{excellent} reliability \cite{ICCGuideline}, supported by a default 95\% confidence interval.

We compute the average of the two separate scores for each project, as shown in Table \ref{tab:humanEval}. It can be seen that our \textit{F2SRD} consistently achieves superior scores in both evaluation criteria across all three projects. To further highlight the distinctions, we conduct Welch's t-test, which substantiates a meaningful disparity between the GPT-4 and \textit{F2SRD} scores at a 95\% confidence interval. 


\begin{table}[htbp]
\small
\caption{Human Scores and t-test Results on SRs of GPT-4 and our F2SRD.}
\begin{center}
\renewcommand{\arraystretch}{1.2}
\begin{tabular}{|c|c|c|c|c|c|}
\hline
\multirow{2}{*}{\textbf{Dataset}} & \multirow{2}{*}{\textbf{Approaches}} & \multicolumn{2}{c|}{\textbf{Specificity}}& \multicolumn{2}{c|}{\textbf{Inspiration}} \\ \cline{3-6}
&&\textbf{score}&\textbf{p-value} & \textbf{score} & \textbf{p-value} \\ \hline
\multirow{2}{*}{ePurse} & GPT-4  &3.37 & \multirow{2}{*}{0.007}&1.63 & \multirow{2}{*}{$10^{-3}$}\\
&F2SRD &\textbf{4.04}& &\textbf{2.27} & \\ \hline
\multirow{2}{*}{CPN} &GPT-4&3.62 & \multirow{2}{*}{$10^{-4}$} &1.70 & \multirow{2}{*}{$10^{-7}$}\\ 
&F2SRD &\textbf{4.08}& & \textbf{2.10} & \\ \hline
\multirow{2}{*}{GPS} & GPT-4 &3.77 & \multirow{2}{*}{$10^{-6}$} &1.58 & \multirow{2}{*}{$10^{-12}$} \\ 
&F2SRD  &\textbf{4.30}& & \textbf{2.32} &  \\ \hline
\end{tabular}
\label{tab:humanEval}
\end{center}
\end{table}

\subsection{Discussion}
\label{subsec:discussion}

According to the results in Tables \ref{tab:quantitativeResults} and \ref{tab:humanEval}, it is clear that our \textit{F2SRD} outperforms the other models across the three key attributes of inspiration, diversity, and specificity. In this section, we will explore the underlying factors contributing to this superiority, as well as the potential failure scenarios.

\textbf{Advantages over RelGAN}. RelGAN aims to generate realistic requirements, but as delineated by its training methodology and technical tenets \cite{koscinski2023demand}, it does not prioritize the generation of ``SRs'' for a ``specific project''. As a result, RelGAN can produce a range of requirements types, not exclusively security-focused, and lacks the capability to target specific projects directionally. In our evaluation, it produced 38.27\% of requirements that were out of scope. Additionally, it currently fails to guarantee syntactic precision consistently, as noted in the study by Koscinski et al. \cite{koscinski2023demand}.

In contrast to RelGAN, we employ the more sophisticated GPT-4 as the foundation of our approach, leveraging its advanced natural language understanding and generation capabilities. Guided by our custom-designed prompt and verification process, our model is adept at recognizing potential vulnerabilities to generate tailored SRs.

\begin{figure*}[!htbp]
	\centering
	\includegraphics[trim = {0 2.5cm 0.5cm 0}, clip, width=\textwidth]{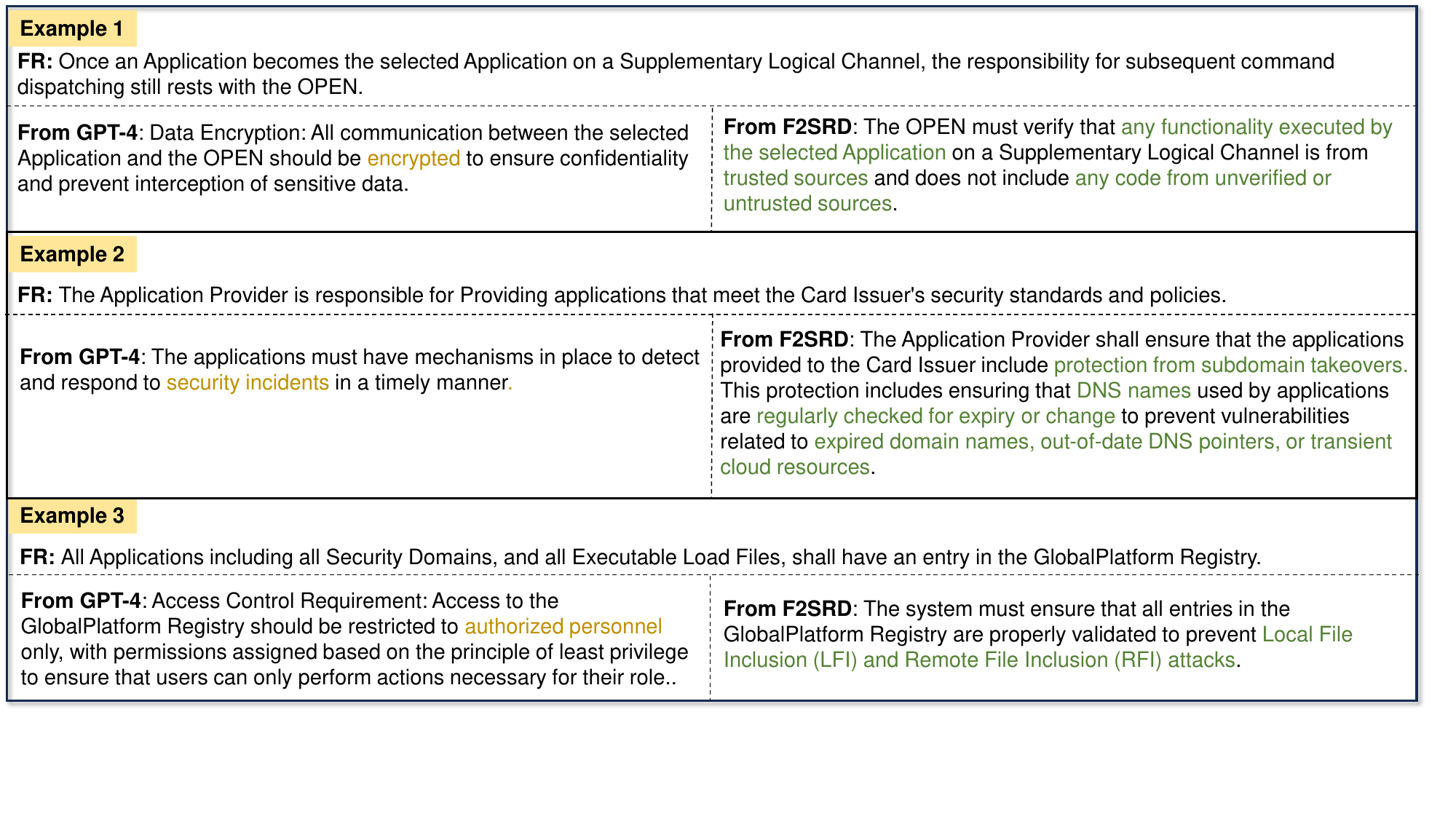}
	\caption{Three Examples showing the Generation of GPT-4 and our \textit{F2SRD}.}
	\label{fig:GPTCompExample}
\end{figure*}

\textbf{Advantages over GPT-4}. To better illustrate the distinction between GPT-4's outputs and those of our \textit{F2SRD}, we present three cases in Fig. \ref{fig:GPTCompExample}. These examples demonstrate that both GPT-4 and \textit{F2SRD} produce SRs that are contextually appropriate and free from syntactical or grammatical errors, surpassing RelGAN's performance.

However, GPT-4's outputs tend to be overly general and may not provide the depth of insight expected in specialized fields. Even junior engineers with basic security knowledge could potentially offer similar responses. By comparison, \textit{F2SRD}'s responses are markedly more detailed. For instance, it specifically recommends protection against \emph{subdomain takeovers} and addresses vulnerabilities associated with \emph{DNS names}—such as the risks involving \emph{expired domain names, outdated DNS pointers, and transient cloud resources}.

The cornerstone of our \textit{F2SRD} lies in the integration of relevant VRs, meticulously crafted from the extensive expertise of numerous security experts across real-world projects. These VRs span a multitude of aspects and phases within application development, ensuring a broad spectrum of consideration. Consequently, this foundation guarantees not only the diversity of the SRs generated by our model but also reinforces their relevance and specificity. 

\textbf{Failure cases analysis.} There are two common failure scenarios.

\textbf{1) Inappropriate VR recommendations leading to erroneous SR production.} The VR recommendations are not always accurate. On one hand, they are constrained by the performance of the retriever; on the other hand, since the retriever processes each FR individually as input, it struggles to handle cross-referencing between FRs. For example, the GPS dataset includes a FR that reads: ``The concept of the Life Cycle of the card (see Section 5.1 - Card Life Cycle) may be used to determine the security level of the communication between the card and an off-card entity.'' The absence of a detailed explanation about the card life cycle in Section 5.1 makes it challenging to fully grasp this FR, thereby increasing the difficulty of interpreting the requirement and retrieving the appropriate VRs.

\begin{figure*}[!htbp]
	\centering
	\includegraphics[trim = {0 13.5cm 0cm 0}, clip, width=\textwidth]{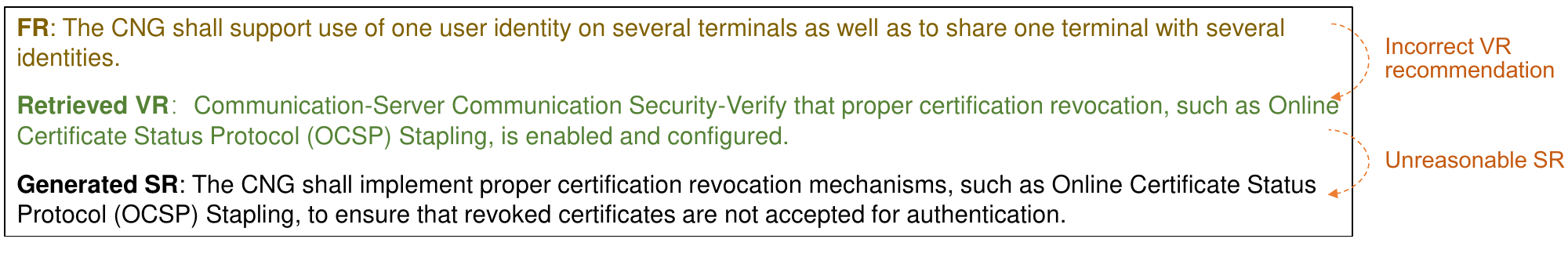}
	\caption{A Failure Case Illustrating An Incorrect VR Recommendation Leads to Inappropriate SR Generation.}
	\label{fig:failureCase}
\end{figure*}

When the suggested VRs do not align with the given FRs, the formulated prompt is deemed unsuitable. Consequently, under these circumstances, deriving SRs that are coherent within the context of the FRs becomes an exceedingly challenging task. For example, as illustrated in Fig. \ref{fig:failureCase}, the FR addresses user account management and the capability to facilitate multiple sessions or users per system. In contrast, the VR focuses on ensuring the security of server-to-server communication channels and the reliability of digital certificates. The former is concerned with identity and session management, while the latter pertains to protocols for secure server communication. Since they pertain to distinct aspects of system security, it is clear that they are not directly correlated. Therefore, it comes as no surprise that the SR generated emphasizes the implementation of a robust certificate revocation mechanism—an aspect that diverges from the intent and scope of the FR.

\begin{figure*}[!htbp]
	\centering
	\includegraphics[trim = {0 11cm 0cm 0}, clip, width=\textwidth]{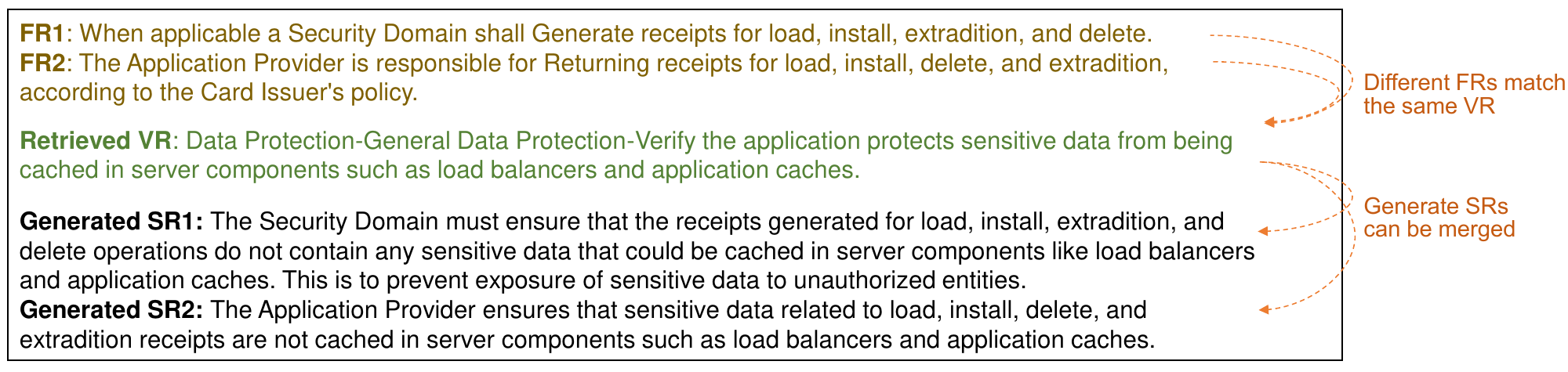}
	\caption{A Failure Case Illustrating some SRs for different FRs can be merged.}
	\label{fig:failureCase2}
\end{figure*}

\noindent \textbf{2) Redundant SRs may be generated.} Since \textit{F2SRD} generates SRs by retrieving VRs independently for each FR, it is possible that different FRs might match the same VR. In such cases, \textit{F2SRD} may produce similar SRs that could actually be merged, as demonstrated in the example in Fig. \ref{fig:failureCase2}. In practical engineering scenarios, it is advisable to consolidate such repetitive SRs into a single, cohesive requirement prior to submission for engineering review, ensuring a streamlined and efficient evaluation process.

\section{Threats to Validity}
\label{sec:discussion}
\textbf{Construct Validity}. The first aspect of construct validity concerns the evaluation criteria and metric selection.
This work aims to provide inspiration and aid to requirements engineers in crafting superior SRs. Consequently, we have adopted inspiration, diversity, and specificity as the focal attributes for assessing the SRs produced. Our quantitative analysis incorporates established metrics such as self-information, self-BLEU, and vocabulary size. Additionally, we have implemented a human-centered qualitative assessment. We are confident that this extensive evaluation framework will more effectively appraise the merits of our approach.


    

Another threat to construct validity arises from the synthesized (FR-VR) pairs used for VR retriever training. We did not assess the quality of this synthetically generated data, which  affects the performance of the VR retriever. To appraise the VR retriever's effectiveness, we carry out an experimental evaluation detailed in Section \ref{sec:RetriverEffectiveness}.

\noindent \textbf{Internal Validity.} The primary threat to internal validity concerns the subjective nature of manual evaluations undertaken for the VR retriever results (see Section \ref{sec:RetriverEffectiveness}) and the generated SRs (see Section \ref{sec:effectiveEva}). Subjectivity, a common challenge in studies requiring human judgment, is unavoidable. To reduce this effect, we employ a two-step process involving independent annotation followed by collective discussion. Moreover, for evaluating the generated SRs, we compute the ICC to quantify annotator agreement. The results indicate high levels of concordance, as detailed in Sections \ref{sec:effectiveResults}.

The second potential issue arises from the filtering of training data. We retain only the FR-VR pairs after filtering by ColBERTv2, with our VR retriever built upon the ColBERTv2 architecture. Consequently, there is a risk that our evaluation results for ColBERTv2 might be biased. To minimize this effect, we have employed an entirely distinct and authentic set of FRs as our testing dataset.

Another factor potentially compromising internal validity arises from the absence of an assessment gauging the `actual' adoption of our generated SRs and their congruence with the ASVS. Currently, our study demonstrates that our methodology can produce inspired, diverse, and specific SRs by weaving VR insights into the generation of SRs. However, additional evaluations are required to establish their practical utility and relevance.

\noindent \textbf{External Validity}. The coupling of our \textit{F2SRD} with the ASVS poses a threat to external validity. In this preliminary study, we opt for ASVS as the strive to generate more useful SRs. Consequently, we synthesize a training dataset and tailor the VR retriever to this specific standard. Although other security standards exist—such as ISO/IEC 15408 (Common Criteria)\cite{valiiso} and FIPS 200—our method theoretically possesses the flexibility to be adapted to these additional standards. However, further evaluations are necessary to substantiate this claim.

\section{Related Work}
\label{sec:relatedWork}

The elicitation and generation of SRs is a process of identifying, analyzing, and specifying the SRs of a system. 
Souag et al. employed a model-driven approach to elicit and generate domain-specific SRs of an information system. This approach takes security goals (e.g., maintain the confidentiality of financial situation) and domain ontologies as inputs, and semi automatically outputs SRs specifications based on a self-defined template \cite{Souag2018AMANDA}. Riaz et al. used machine learning techniques to identify implicit security-related sentences within natural language requirements artifacts (requirement specifications, feature requests, etc.) and associated security goals \cite{6912260}. They also provided context-specific templates (e.g., the system shall log every time <subject> <action> <on|for> <resource>) to assist in converting security-related sentences into SRs. Pandey et al. proposed the SR extraction method STORE \cite{ANSARI2022191}, which is based on security threat analysis. It prioritizes threats by analyzing potential security attacks on the system, aiding SRs engineers in manually eliciting SRs through access to threat dictionaries. Koscinski et al. investigated the use of Relational Generative Adversarial Networks (GANs) for the automated synthesis of SRs specifications \cite{10172729}, demonstrating positive results in the judicial domain. Ahmad et al. proposed a fuzzy-based approach for prioritizing FRs from a security perspective, allowing that requirements engineers prioritize extracting SRs from FRs with higher security relevance. Li et al. proposed a framework that takes information system’s requirements specification as input and uses a formalized vulnerability detection model to automatically detect system vulnerabilities. The framework maps these vulnerabilities to corresponding threats and derives the system's security objectives, ultimately leading to the manual generation of SRs documentation \cite{8009939}. While these approaches are helpful for SRs elicitation, most of them do not achieve full automation in generating SRs, resulting in significant resource and time consumption. Additionally, excessive reliance on templates may lead to SRs that are imprecise or lack specificity.

Besides SRs generation, lots of research focus on general requirements elicitation and generation. Zhao et al. \cite{ReqGen} generated requirement drafts by integrating domain ontology with the Unified pre-trained Language Model (UniLM), although their method may be hindered by the often-absent domain ontologies. Gudaparthi et al.\cite{10260978} concentrate on creativity within RE and suggest the generation of creative requirements using adversarial examples. Model-driven requirements engineering is another significant approach, which typically converts structured models like business process models \cite{turetken2004automating, DBLP:journals/infsof/CoxPBV05}, the i* framework \cite{DBLP:journals/re/MaidenMJG05, DBLP:conf/coopis/YuBDM95}, KAOS and Objectiver \cite{DBLP:conf/sigsoft/LetierL02, DBLP:journals/re/LandtsheerLL04, van2004goal}, and UML models \cite{DBLP:journals/tse/LamsweerdeW98, DBLP:journals/re/MezianeAA08, DBLP:conf/re/Berenbach03a,9793770} into structured natural language specifications using set rules. The precision of input models and the limitations imposed by these rules, however, require substantial human input and restrict flexibility.

\section{Conclusion}
\label{sec:conclusion}

We introduce \textit{F2SRD}, a novel methodology for the automated generation of inspired, diverse, and specific SRs from given FRs, guided by relevant security verification criteria. To facilitate this process, we have developed and trained a VR retriever to adeptly identify pertinent VRs from the widely-accepted ASVS, corresponding to the supplied FRs. Faced with the scarcity of publicly accessible FR-VR pairs, we employed GPT-4 to synthesize these pairs, the efficacy of which is validated through our experiments (detailed in Section \ref{sec:RetriverEffectiveness}). Subsequently, we utilize both the original FRs and the relevant VRs retrieved to create targeted prompts that steer GPT-4 in generating high-quality SRs. Both quantitative and qualitative analyses demonstrate that our approach significantly surpasses benchmarks set by RelGAN and standalone GPT-4 models in producing inspired, diverse, and specific SRs. Our approach has the potential to be used in security-critical systems to enhance the completeness of SRs and help improve the overall security of the final software systems. Looking ahead, we aim to evaluate the practical applicability of our generated SRs in real-world scenarios and their congruence with ASVS standards through further experimentation.

\section{DATA AVAILABILITY}
\changed{We provide the data and scripts online to facilitate replications or future work: \url{https://doi.org/10.6084/m9.figshare.27014170.v6}}.

\section*{ACKNOWLEDGMENTS}
We thank the associated editors and reviewers of FSE 2025 for their valuable feedback. Funding for this work was provided by the National Natural Science Foundation of China (Grants No. 62102014 and 62177003) and the National Science and Technology Major Project of China (Grant No. Y2022-V-0001-0027).

\bibliographystyle{ACM-Reference-Format}
\bibliography{Reference}


\begin{thebibliography}{59}


\ifx \showCODEN    \undefined \def \showCODEN     #1{\unskip}     \fi
\ifx \showDOI      \undefined \def \showDOI       #1{#1}\fi
\ifx \showISBNx    \undefined \def \showISBNx     #1{\unskip}     \fi
\ifx \showISBNxiii \undefined \def \showISBNxiii  #1{\unskip}     \fi
\ifx \showISSN     \undefined \def \showISSN      #1{\unskip}     \fi
\ifx \showLCCN     \undefined \def \showLCCN      #1{\unskip}     \fi
\ifx \shownote     \undefined \def \shownote      #1{#1}          \fi
\ifx \showarticletitle \undefined \def \showarticletitle #1{#1}   \fi
\ifx \showURL      \undefined \def \showURL       {\relax}        \fi
\providecommand\bibfield[2]{#2}
\providecommand\bibinfo[2]{#2}
\providecommand\natexlab[1]{#1}
\providecommand\showeprint[2][]{arXiv:#2}

\bibitem[846(2018)]%
        {8463987}
 \bibinfo{year}{2018}\natexlab{}.
\newblock \showarticletitle{ISO/IEC/IEEE Draft International Standard - Systems and Software Engineering -- Life Cycle Processes --Requirements Engineering}.
\newblock \bibinfo{journal}{\emph{ISO/IEC/IEEE P29148\_FDIS, September 2018}} (\bibinfo{year}{2018}), \bibinfo{pages}{1--104}.
\newblock


\bibitem[Achiam et~al\mbox{.}(2023)]%
        {achiam2023gpt}
\bibfield{author}{\bibinfo{person}{Josh Achiam}, \bibinfo{person}{Steven Adler}, \bibinfo{person}{Sandhini Agarwal}, \bibinfo{person}{Lama Ahmad}, \bibinfo{person}{Ilge Akkaya}, \bibinfo{person}{Florencia~Leoni Aleman}, \bibinfo{person}{Diogo Almeida}, \bibinfo{person}{Janko Altenschmidt}, \bibinfo{person}{Sam Altman}, \bibinfo{person}{Shyamal Anadkat}, {et~al\mbox{.}}} \bibinfo{year}{2023}\natexlab{}.
\newblock \showarticletitle{Gpt-4 technical report}.
\newblock \bibinfo{journal}{\emph{arXiv preprint arXiv:2303.08774}} (\bibinfo{year}{2023}).
\newblock


\bibitem[Ahmad et~al\mbox{.}(2021)]%
        {9397153}
\bibfield{author}{\bibinfo{person}{Javed Ahmad}, \bibinfo{person}{Chaudhary~Wali Mohammad}, {and} \bibinfo{person}{Mohd Sadiq}.} \bibinfo{year}{2021}\natexlab{}.
\newblock \showarticletitle{Identification of Security Requirements from the Selected Set of Requirements under Fuzzy Environment}. In \bibinfo{booktitle}{\emph{2021 International Conference on Computing, Communication, and Intelligent Systems (ICCCIS)}}. \bibinfo{pages}{58--63}.
\newblock
\urldef\tempurl%
\url{https://doi.org/10.1109/ICCCIS51004.2021.9397153}
\showDOI{\tempurl}


\bibitem[Ansari and Pandey(2018)]%
        {inbook}
\bibfield{author}{\bibinfo{person}{Md~Tarique Ansari} {and} \bibinfo{person}{Dhirendra Pandey}.} \bibinfo{year}{2018}\natexlab{}.
\newblock \bibinfo{booktitle}{\emph{Risks, security, and privacy for HIV/AIDS data: Big data perspective}}.
\newblock \bibinfo{pages}{117--139}.
\newblock
\showISBNx{9781522532040}
\urldef\tempurl%
\url{https://doi.org/10.4018/978-1-5225-3203-3.ch005}
\showDOI{\tempurl}


\bibitem[Ansari et~al\mbox{.}(2022)]%
        {ANSARI2022191}
\bibfield{author}{\bibinfo{person}{Md~Tarique~Jamal Ansari}, \bibinfo{person}{Dhirendra Pandey}, {and} \bibinfo{person}{Mamdouh Alenezi}.} \bibinfo{year}{2022}\natexlab{}.
\newblock \showarticletitle{STORE: Security Threat Oriented Requirements Engineering Methodology}.
\newblock \bibinfo{journal}{\emph{Journal of King Saud University - Computer and Information Sciences}} \bibinfo{volume}{34}, \bibinfo{number}{2} (\bibinfo{year}{2022}), \bibinfo{pages}{191--203}.
\newblock
\showISSN{1319-1578}
\urldef\tempurl%
\url{https://doi.org/10.1016/j.jksuci.2018.12.005}
\showDOI{\tempurl}


\bibitem[Bao et~al\mbox{.}(2022)]%
        {9793770}
\bibfield{author}{\bibinfo{person}{Tianshu Bao}, \bibinfo{person}{Jing Yang}, \bibinfo{person}{Yilong Yang}, {and} \bibinfo{person}{Yongfeng Yin}.} \bibinfo{year}{2022}\natexlab{}.
\newblock \showarticletitle{RM2Doc: A Tool for Automatic Generation of Requirements Documents from Requirements Models}. In \bibinfo{booktitle}{\emph{2022 IEEE/ACM 44th International Conference on Software Engineering: Companion Proceedings (ICSE-Companion)}}. \bibinfo{pages}{188--192}.
\newblock
\urldef\tempurl%
\url{https://doi.org/10.1145/3510454.3516850}
\showDOI{\tempurl}


\bibitem[Berenbach(2003)]%
        {DBLP:conf/re/Berenbach03a}
\bibfield{author}{\bibinfo{person}{Brian Berenbach}.} \bibinfo{year}{2003}\natexlab{}.
\newblock \showarticletitle{The Automated Extraction of Requirements from {UML} Models}. In \bibinfo{booktitle}{\emph{11th {IEEE} International Conference on Requirements Engineering {(RE} 2003), 8-12 September 2003, Monterey Bay, CA, {USA}}}. \bibinfo{publisher}{{IEEE} Computer Society}, \bibinfo{pages}{287}.
\newblock


\bibitem[Bonifacio et~al\mbox{.}(2022)]%
        {bonifacio2022inpars}
\bibfield{author}{\bibinfo{person}{Luiz Bonifacio}, \bibinfo{person}{Hugo Abonizio}, \bibinfo{person}{Marzieh Fadaee}, {and} \bibinfo{person}{Rodrigo Nogueira}.} \bibinfo{year}{2022}\natexlab{}.
\newblock \showarticletitle{Inpars: Data augmentation for information retrieval using large language models}.
\newblock \bibinfo{journal}{\emph{arXiv preprint arXiv:2202.05144}} (\bibinfo{year}{2022}).
\newblock


\bibitem[Bulusu et~al\mbox{.}(2017)]%
        {bulusu2017towards}
\bibfield{author}{\bibinfo{person}{Sravani~Teja Bulusu}, \bibinfo{person}{Romain Laborde}, \bibinfo{person}{Ahmad~Samer Wazan}, \bibinfo{person}{Fran{\c{c}}ois Barr{\`e}re}, {and} \bibinfo{person}{Abdelmalek Benzekri}.} \bibinfo{year}{2017}\natexlab{}.
\newblock \showarticletitle{Towards the weaving of the characteristics of good security requirements}. In \bibinfo{booktitle}{\emph{Risks and Security of Internet and Systems: 11th International Conference, CRiSIS 2016, Roscoff, France, September 5-7, 2016, Revised Selected Papers 11}}. Springer, \bibinfo{pages}{60--74}.
\newblock


\bibitem[Cox et~al\mbox{.}(2005)]%
        {DBLP:journals/infsof/CoxPBV05}
\bibfield{author}{\bibinfo{person}{Karl Cox}, \bibinfo{person}{Keith Phalp}, \bibinfo{person}{Steven~J. Bleistein}, {and} \bibinfo{person}{June~M. Verner}.} \bibinfo{year}{2005}\natexlab{}.
\newblock \showarticletitle{Deriving requirements from process models via the problem frames approach}.
\newblock \bibinfo{journal}{\emph{Inf. Softw. Technol.}} \bibinfo{volume}{47}, \bibinfo{number}{5} (\bibinfo{year}{2005}), \bibinfo{pages}{319--337}.
\newblock


\bibitem[Dai et~al\mbox{.}(2022)]%
        {dai2022promptagator}
\bibfield{author}{\bibinfo{person}{Zhuyun Dai}, \bibinfo{person}{Vincent~Y Zhao}, \bibinfo{person}{Ji Ma}, \bibinfo{person}{Yi Luan}, \bibinfo{person}{Jianmo Ni}, \bibinfo{person}{Jing Lu}, \bibinfo{person}{Anton Bakalov}, \bibinfo{person}{Kelvin Guu}, \bibinfo{person}{Keith~B Hall}, {and} \bibinfo{person}{Ming-Wei Chang}.} \bibinfo{year}{2022}\natexlab{}.
\newblock \showarticletitle{PROMPTAGATOR: FEW-SHOT DENSE RETRIEVAL FROM 8 EXAMPLES}.
\newblock \bibinfo{journal}{\emph{arXiv preprint arXiv:2209.11755}} (\bibinfo{year}{2022}).
\newblock


\bibitem[Dubey et~al\mbox{.}(2024)]%
        {dubey2024llama}
\bibfield{author}{\bibinfo{person}{Abhimanyu Dubey}, \bibinfo{person}{Abhinav Jauhri}, \bibinfo{person}{Abhinav Pandey}, \bibinfo{person}{Abhishek Kadian}, \bibinfo{person}{Ahmad Al-Dahle}, \bibinfo{person}{Aiesha Letman}, \bibinfo{person}{Akhil Mathur}, \bibinfo{person}{Alan Schelten}, \bibinfo{person}{Amy Yang}, \bibinfo{person}{Angela Fan}, {et~al\mbox{.}}} \bibinfo{year}{2024}\natexlab{}.
\newblock \showarticletitle{The llama 3 herd of models}.
\newblock \bibinfo{journal}{\emph{arXiv preprint arXiv:2407.21783}} (\bibinfo{year}{2024}).
\newblock


\bibitem[Elder et~al\mbox{.}(2021)]%
        {9402211}
\bibfield{author}{\bibinfo{person}{Sarah Elder}, \bibinfo{person}{Nusrat Zahan}, \bibinfo{person}{Valeri Kozarev}, \bibinfo{person}{Rui Shu}, \bibinfo{person}{Tim Menzies}, {and} \bibinfo{person}{Laurie Williams}.} \bibinfo{year}{2021}\natexlab{}.
\newblock \showarticletitle{Structuring a Comprehensive Software Security Course Around the OWASP Application Security Verification Standard}. In \bibinfo{booktitle}{\emph{2021 IEEE/ACM 43rd International Conference on Software Engineering: Software Engineering Education and Training (ICSE-SEET)}}. \bibinfo{pages}{95--104}.
\newblock
\urldef\tempurl%
\url{https://doi.org/10.1109/ICSE-SEET52601.2021.00019}
\showDOI{\tempurl}


\bibitem[Fern\'{a}ndez et~al\mbox{.}(2017)]%
        {10.1007/s10664-016-9451-7}
\bibfield{author}{\bibinfo{person}{D.~M\'{e}ndez Fern\'{a}ndez}, \bibinfo{person}{S. Wagner}, \bibinfo{person}{M. Kalinowski}, \bibinfo{person}{M. Felderer}, \bibinfo{person}{P. Mafra}, \bibinfo{person}{A. Vetr\`{o}}, \bibinfo{person}{T. Conte}, \bibinfo{person}{M.~T. Christiansson}, \bibinfo{person}{D. Greer}, \bibinfo{person}{C. Lassenius}, \bibinfo{person}{T. M\"{a}nnist\"{o}}, \bibinfo{person}{M. Nayabi}, \bibinfo{person}{M. Oivo}, \bibinfo{person}{B. Penzenstadler}, \bibinfo{person}{D. Pfahl}, \bibinfo{person}{R. Prikladnicki}, \bibinfo{person}{G. Ruhe}, \bibinfo{person}{A. Schekelmann}, \bibinfo{person}{S. Sen}, \bibinfo{person}{R. Spinola}, \bibinfo{person}{A. Tuzcu}, \bibinfo{person}{J.~L. De~La~Vara}, {and} \bibinfo{person}{R. Wieringa}.} \bibinfo{year}{2017}\natexlab{}.
\newblock \showarticletitle{Naming the pain in requirements engineering}.
\newblock \bibinfo{journal}{\emph{Empirical Softw. Eng.}} \bibinfo{volume}{22}, \bibinfo{number}{5} (\bibinfo{date}{oct} \bibinfo{year}{2017}), \bibinfo{pages}{2298–2338}.
\newblock
\showISSN{1382-3256}
\urldef\tempurl%
\url{https://doi.org/10.1007/s10664-016-9451-7}
\showDOI{\tempurl}


\bibitem[Fisher(1992)]%
        {Fisher1992}
\bibfield{author}{\bibinfo{person}{R.~A. Fisher}.} \bibinfo{year}{1992}\natexlab{}.
\newblock \bibinfo{booktitle}{\emph{Statistical Methods for Research Workers}}.
\newblock \bibinfo{publisher}{Springer New York}, \bibinfo{address}{New York, NY}, \bibinfo{pages}{66--70}.
\newblock
\showISBNx{978-1-4612-4380-9}
\urldef\tempurl%
\url{https://doi.org/10.1007/978-1-4612-4380-9_6}
\showDOI{\tempurl}


\bibitem[Fong(2018)]%
        {fong2018software}
\bibfield{author}{\bibinfo{person}{Vivian Fong}.} \bibinfo{year}{2018}\natexlab{}.
\newblock \emph{\bibinfo{title}{Software requirements classification using word embeddings and convolutional neural networks}}.
\newblock \bibinfo{thesistype}{Master's\ thesis}. \bibinfo{school}{California Polytechnic State University}.
\newblock


\bibitem[Gudaparthi et~al\mbox{.}(2023)]%
        {10260978}
\bibfield{author}{\bibinfo{person}{Hemanth Gudaparthi}, \bibinfo{person}{Nan Niu}, \bibinfo{person}{Boyang Wang}, \bibinfo{person}{Tanmay Bhowmik}, \bibinfo{person}{Hui Liu}, \bibinfo{person}{Jianzhang Zhang}, \bibinfo{person}{Juha Savolainen}, \bibinfo{person}{Glen Horton}, \bibinfo{person}{Sean Crowe}, \bibinfo{person}{Thomas Scherz}, {and} \bibinfo{person}{Lisa Haitz}.} \bibinfo{year}{2023}\natexlab{}.
\newblock \showarticletitle{Prompting Creative Requirements via Traceable and Adversarial Examples in Deep Learning}. In \bibinfo{booktitle}{\emph{2023 IEEE 31st International Requirements Engineering Conference (RE)}}. \bibinfo{pages}{134--145}.
\newblock
\urldef\tempurl%
\url{https://doi.org/10.1109/RE57278.2023.00022}
\showDOI{\tempurl}


\bibitem[Houmb et~al\mbox{.}(2010)]%
        {houmb2010eliciting}
\bibfield{author}{\bibinfo{person}{Siv~Hilde Houmb}, \bibinfo{person}{Shareeful Islam}, \bibinfo{person}{Eric Knauss}, \bibinfo{person}{Jan J{\"u}rjens}, {and} \bibinfo{person}{Kurt Schneider}.} \bibinfo{year}{2010}\natexlab{}.
\newblock \showarticletitle{Eliciting security requirements and tracing them to design: an integration of Common Criteria, heuristics, and UMLsec}.
\newblock \bibinfo{journal}{\emph{Requirements Engineering}}  \bibinfo{volume}{15} (\bibinfo{year}{2010}), \bibinfo{pages}{63--93}.
\newblock


\bibitem[ISO(2011)]%
        {iso2011ieee}
\bibfield{author}{\bibinfo{person}{IEC ISO}.} \bibinfo{year}{2011}\natexlab{}.
\newblock \showarticletitle{Ieee: Iso/iec/ieee 29148, systems and software engineering, life cycle processes}.
\newblock \bibinfo{journal}{\emph{Requirements engineering}}  \bibinfo{volume}{600} (\bibinfo{year}{2011}).
\newblock


\bibitem[Jeronymo et~al\mbox{.}(2023)]%
        {jeronymo2023inpars}
\bibfield{author}{\bibinfo{person}{Vitor Jeronymo}, \bibinfo{person}{Luiz Bonifacio}, \bibinfo{person}{Hugo Abonizio}, \bibinfo{person}{Marzieh Fadaee}, \bibinfo{person}{Roberto Lotufo}, \bibinfo{person}{Jakub Zavrel}, {and} \bibinfo{person}{Rodrigo Nogueira}.} \bibinfo{year}{2023}\natexlab{}.
\newblock \showarticletitle{Inpars-v2: Large language models as efficient dataset generators for information retrieval}.
\newblock \bibinfo{journal}{\emph{arXiv preprint arXiv:2301.01820}} (\bibinfo{year}{2023}).
\newblock


\bibitem[Katona and Nemetz(1976)]%
        {katona1976huffman}
\bibfield{author}{\bibinfo{person}{Gyula Katona} {and} \bibinfo{person}{O Nemetz}.} \bibinfo{year}{1976}\natexlab{}.
\newblock \showarticletitle{Huffman codes and self-information}.
\newblock \bibinfo{journal}{\emph{IEEE Transactions on Information Theory}} \bibinfo{volume}{22}, \bibinfo{number}{3} (\bibinfo{year}{1976}), \bibinfo{pages}{337--340}.
\newblock


\bibitem[Knauss et~al\mbox{.}(2011)]%
        {10.1007/978-3-642-19858-8_2}
\bibfield{author}{\bibinfo{person}{Eric Knauss}, \bibinfo{person}{Siv Houmb}, \bibinfo{person}{Kurt Schneider}, \bibinfo{person}{Shareeful Islam}, {and} \bibinfo{person}{Jan J{\"u}rjens}.} \bibinfo{year}{2011}\natexlab{}.
\newblock \showarticletitle{Supporting Requirements Engineers in Recognising Security Issues}. In \bibinfo{booktitle}{\emph{Requirements Engineering: Foundation for Software Quality}}, \bibfield{editor}{\bibinfo{person}{Daniel Berry} {and} \bibinfo{person}{Xavier Franch}} (Eds.). \bibinfo{publisher}{Springer Berlin Heidelberg}, \bibinfo{address}{Berlin, Heidelberg}, \bibinfo{pages}{4--18}.
\newblock
\showISBNx{978-3-642-19858-8}


\bibitem[Koo and Li(2016)]%
        {ICCGuideline}
\bibfield{author}{\bibinfo{person}{Terry~K Koo} {and} \bibinfo{person}{Mae~Y Li}.} \bibinfo{year}{2016}\natexlab{}.
\newblock \showarticletitle{A Guideline of Selecting and Reporting Intraclass Correlation Coefficients for Reliability Research}.
\newblock \bibinfo{journal}{\emph{Journal of chiropractic medicine}}  \bibinfo{volume}{15} (\bibinfo{year}{2016}), \bibinfo{pages}{155--163}.
\newblock
Issue 2.


\bibitem[Koscinski et~al\mbox{.}(2023a)]%
        {koscinski2023demand}
\bibfield{author}{\bibinfo{person}{Viktoria Koscinski}, \bibinfo{person}{Sara Hashemi}, {and} \bibinfo{person}{Mehdi Mirakhorli}.} \bibinfo{year}{2023}\natexlab{a}.
\newblock \showarticletitle{On-demand security requirements synthesis with relational generative adversarial networks}. In \bibinfo{booktitle}{\emph{2023 IEEE/ACM 45th International Conference on Software Engineering (ICSE)}}. IEEE, \bibinfo{pages}{1609--1621}.
\newblock


\bibitem[Koscinski et~al\mbox{.}(2023b)]%
        {10172729}
\bibfield{author}{\bibinfo{person}{Viktoria Koscinski}, \bibinfo{person}{Sara Hashemi}, {and} \bibinfo{person}{Mehdi Mirakhorli}.} \bibinfo{year}{2023}\natexlab{b}.
\newblock \showarticletitle{On-Demand Security Requirements Synthesis with Relational Generative Adversarial Networks}. In \bibinfo{booktitle}{\emph{2023 IEEE/ACM 45th International Conference on Software Engineering (ICSE)}}. \bibinfo{pages}{1609--1621}.
\newblock
\urldef\tempurl%
\url{https://doi.org/10.1109/ICSE48619.2023.00139}
\showDOI{\tempurl}


\bibitem[Krasner(2021)]%
        {CPSQReport}
\bibfield{author}{\bibinfo{person}{Kerb Krasner}.} \bibinfo{year}{2021}\natexlab{}.
\newblock \bibinfo{booktitle}{\emph{The Cost of Poor Software Quality in the US: A 2020 Report}}.
\newblock \bibinfo{type}{{T}echnical {R}eport}. \bibinfo{institution}{CISQ}.
\newblock


\bibitem[Lamsweerde(2009)]%
        {lamsweerde2009requirements}
\bibfield{author}{\bibinfo{person}{A~van Lamsweerde}.} \bibinfo{year}{2009}\natexlab{}.
\newblock \bibinfo{booktitle}{\emph{Requirements engineering: from system goals to UML models to software specifications}}.
\newblock \bibinfo{publisher}{John Wiley \& Sons, Ltd}.
\newblock


\bibitem[Landtsheer et~al\mbox{.}(2004)]%
        {DBLP:journals/re/LandtsheerLL04}
\bibfield{author}{\bibinfo{person}{Renaud~De Landtsheer}, \bibinfo{person}{Emmanuel Letier}, {and} \bibinfo{person}{Axel van Lamsweerde}.} \bibinfo{year}{2004}\natexlab{}.
\newblock \showarticletitle{Deriving tabular event-based specifications from goal-oriented requirements models}.
\newblock \bibinfo{journal}{\emph{Requir. Eng.}} \bibinfo{volume}{9}, \bibinfo{number}{2} (\bibinfo{year}{2004}), \bibinfo{pages}{104--120}.
\newblock


\bibitem[Len~Bass(2021)]%
        {SoftwareArchitecture}
\bibfield{author}{\bibinfo{person}{Rick~Kazman Len~Bass, Paul~Clements}.} \bibinfo{year}{2021}\natexlab{}.
\newblock \bibinfo{booktitle}{\emph{Software Architecture in Practice, 4th Edition}}.
\newblock \bibinfo{publisher}{Addison-Wesley Professional}.
\newblock


\bibitem[Letier and van Lamsweerde(2002)]%
        {DBLP:conf/sigsoft/LetierL02}
\bibfield{author}{\bibinfo{person}{Emmanuel Letier} {and} \bibinfo{person}{Axel van Lamsweerde}.} \bibinfo{year}{2002}\natexlab{}.
\newblock \showarticletitle{Deriving operational software specifications from system goals}. In \bibinfo{booktitle}{\emph{Proceedings of the Tenth {ACM} {SIGSOFT} Symposium on Foundations of Software Engineering 2002, Charleston, South Carolina, USA, November 18-22, 2002}}. \bibinfo{publisher}{{ACM}}, \bibinfo{pages}{119--128}.
\newblock


\bibitem[Li et~al\mbox{.}(2017)]%
        {8009939}
\bibfield{author}{\bibinfo{person}{Hongbo Li}, \bibinfo{person}{Xiaohong Li}, \bibinfo{person}{Jianye Hao}, \bibinfo{person}{Guangquan Xu}, \bibinfo{person}{Zhiyong Feng}, {and} \bibinfo{person}{Xiaofei Xie}.} \bibinfo{year}{2017}\natexlab{}.
\newblock \showarticletitle{FESR: A Framework for Eliciting Security Requirements Based on Integration of Common Criteria and Weakness Detection Formal Model}. In \bibinfo{booktitle}{\emph{2017 IEEE International Conference on Software Quality, Reliability and Security (QRS)}}. \bibinfo{pages}{352--363}.
\newblock
\urldef\tempurl%
\url{https://doi.org/10.1109/QRS.2017.45}
\showDOI{\tempurl}


\bibitem[Maiden et~al\mbox{.}(2005)]%
        {DBLP:journals/re/MaidenMJG05}
\bibfield{author}{\bibinfo{person}{Neil A.~M. Maiden}, \bibinfo{person}{Sharon Manning}, \bibinfo{person}{Sara Jones}, {and} \bibinfo{person}{John Greenwood}.} \bibinfo{year}{2005}\natexlab{}.
\newblock \showarticletitle{Generating requirements from systems models using patterns: a case study}.
\newblock \bibinfo{journal}{\emph{Requir. Eng.}} \bibinfo{volume}{10}, \bibinfo{number}{4} (\bibinfo{year}{2005}), \bibinfo{pages}{276--288}.
\newblock


\bibitem[Markowsky(2024)]%
        {informationTheory}
\bibfield{author}{\bibinfo{person}{George Markowsky}.} \bibinfo{year}{2024}\natexlab{}.
\newblock \showarticletitle{Information Theory}.
\newblock \bibinfo{howpublished}{\url{https://www.britannica.com/science/information-theory}}.
\newblock \bibinfo{journal}{\emph{Encyclopedia Britannica}} (\bibinfo{year}{2024}).
\newblock


\bibitem[Mead and Stehney(2005)]%
        {10.1145/1082983.1083214}
\bibfield{author}{\bibinfo{person}{Nancy~R. Mead} {and} \bibinfo{person}{Ted Stehney}.} \bibinfo{year}{2005}\natexlab{}.
\newblock \showarticletitle{Security quality requirements engineering (SQUARE) methodology}.
\newblock \bibinfo{journal}{\emph{SIGSOFT Softw. Eng. Notes}} \bibinfo{volume}{30}, \bibinfo{number}{4} (\bibinfo{date}{may} \bibinfo{year}{2005}), \bibinfo{pages}{1–7}.
\newblock
\showISSN{0163-5948}
\urldef\tempurl%
\url{https://doi.org/10.1145/1082983.1083214}
\showDOI{\tempurl}


\bibitem[Meziane et~al\mbox{.}(2008)]%
        {DBLP:journals/re/MezianeAA08}
\bibfield{author}{\bibinfo{person}{Farid Meziane}, \bibinfo{person}{Nikos Athanasakis}, {and} \bibinfo{person}{Sophia Ananiadou}.} \bibinfo{year}{2008}\natexlab{}.
\newblock \showarticletitle{Generating Natural Language specifications from {UML} class diagrams}.
\newblock \bibinfo{journal}{\emph{Requir. Eng.}} \bibinfo{volume}{13}, \bibinfo{number}{1} (\bibinfo{year}{2008}), \bibinfo{pages}{1--18}.
\newblock


\bibitem[Nanisura~Damanik and Sunaringtyas(2020)]%
        {9255559}
\bibfield{author}{\bibinfo{person}{Venia~Noella Nanisura~Damanik} {and} \bibinfo{person}{Septia~Ulfa Sunaringtyas}.} \bibinfo{year}{2020}\natexlab{}.
\newblock \showarticletitle{Secure Code Recommendation Based on Code Review Result Using OWASP Code Review Guide}. In \bibinfo{booktitle}{\emph{2020 International Workshop on Big Data and Information Security (IWBIS)}}. \bibinfo{pages}{153--158}.
\newblock
\urldef\tempurl%
\url{https://doi.org/10.1109/IWBIS50925.2020.9255559}
\showDOI{\tempurl}


\bibitem[Nie et~al\mbox{.}(2019)]%
        {nie2018relgan}
\bibfield{author}{\bibinfo{person}{Weili Nie}, \bibinfo{person}{Nina Narodytska}, {and} \bibinfo{person}{Ankit Patel}.} \bibinfo{year}{2019}\natexlab{}.
\newblock \showarticletitle{Rel{GAN}: Relational Generative Adversarial Networks for Text Generation}. In \bibinfo{booktitle}{\emph{International Conference on Learning Representations}}.
\newblock
\urldef\tempurl%
\url{https://openreview.net/forum?id=rJedV3R5tm}
\showURL{%
\tempurl}


\bibitem[Riaz et~al\mbox{.}(2014)]%
        {6912260}
\bibfield{author}{\bibinfo{person}{Maria Riaz}, \bibinfo{person}{Jason King}, \bibinfo{person}{John Slankas}, {and} \bibinfo{person}{Laurie Williams}.} \bibinfo{year}{2014}\natexlab{}.
\newblock \showarticletitle{Hidden in plain sight: Automatically identifying security requirements from natural language artifacts}. In \bibinfo{booktitle}{\emph{2014 IEEE 22nd International Requirements Engineering Conference (RE)}}. \bibinfo{pages}{183--192}.
\newblock
\urldef\tempurl%
\url{https://doi.org/10.1109/RE.2014.6912260}
\showDOI{\tempurl}


\bibitem[Saad-Falcon et~al\mbox{.}(2023)]%
        {saad-falcon-etal-2023-udapdr}
\bibfield{author}{\bibinfo{person}{Jon Saad-Falcon}, \bibinfo{person}{Omar Khattab}, \bibinfo{person}{Keshav Santhanam}, \bibinfo{person}{Radu Florian}, \bibinfo{person}{Martin Franz}, \bibinfo{person}{Salim Roukos}, \bibinfo{person}{Avirup Sil}, \bibinfo{person}{Md Sultan}, {and} \bibinfo{person}{Christopher Potts}.} \bibinfo{year}{2023}\natexlab{}.
\newblock \showarticletitle{{UDAPDR}: Unsupervised Domain Adaptation via {LLM} Prompting and Distillation of Rerankers}. In \bibinfo{booktitle}{\emph{Proceedings of the 2023 Conference on Empirical Methods in Natural Language Processing}}, \bibfield{editor}{\bibinfo{person}{Houda Bouamor}, \bibinfo{person}{Juan Pino}, {and} \bibinfo{person}{Kalika Bali}} (Eds.). \bibinfo{publisher}{Association for Computational Linguistics}, \bibinfo{address}{Singapore}, \bibinfo{pages}{11265--11279}.
\newblock
\urldef\tempurl%
\url{https://doi.org/10.18653/v1/2023.emnlp-main.693}
\showDOI{\tempurl}


\bibitem[Santhanam et~al\mbox{.}(2021a)]%
        {santhanam2021colbertv2}
\bibfield{author}{\bibinfo{person}{Keshav Santhanam}, \bibinfo{person}{Omar Khattab}, \bibinfo{person}{Jon Saad-Falcon}, \bibinfo{person}{Christopher Potts}, {and} \bibinfo{person}{Matei Zaharia}.} \bibinfo{year}{2021}\natexlab{a}.
\newblock \showarticletitle{Colbertv2: Effective and efficient retrieval via lightweight late interaction}.
\newblock \bibinfo{journal}{\emph{arXiv preprint arXiv:2112.01488}} (\bibinfo{year}{2021}).
\newblock


\bibitem[Santhanam et~al\mbox{.}(2021b)]%
        {Santhanam2021ColBERTv2EA}
\bibfield{author}{\bibinfo{person}{Keshav Santhanam}, \bibinfo{person}{O. Khattab}, \bibinfo{person}{Jon Saad-Falcon}, \bibinfo{person}{Christopher Potts}, {and} \bibinfo{person}{Matei~A. Zaharia}.} \bibinfo{year}{2021}\natexlab{b}.
\newblock \showarticletitle{ColBERTv2: Effective and Efficient Retrieval via Lightweight Late Interaction}. In \bibinfo{booktitle}{\emph{North American Chapter of the Association for Computational Linguistics}}.
\newblock
\urldef\tempurl%
\url{https://api.semanticscholar.org/CorpusID:244799249}
\showURL{%
\tempurl}


\bibitem[Security(2020)]%
        {contrastSecurity}
\bibfield{author}{\bibinfo{person}{Contrast Security}.} \bibinfo{year}{2020}\natexlab{}.
\newblock \bibinfo{title}{2020 Application Security Observability Report}.
\newblock \bibinfo{howpublished}{\url{https://www.contrastsecurity.com/hubfs/2020-Contrast-Labs-Application-Security-Observability_Annual_Report_07152020.pdf}}.
\newblock


\bibitem[Shannon(1948)]%
        {6773024}
\bibfield{author}{\bibinfo{person}{C.~E. Shannon}.} \bibinfo{year}{1948}\natexlab{}.
\newblock \showarticletitle{A mathematical theory of communication}.
\newblock \bibinfo{journal}{\emph{The Bell System Technical Journal}} \bibinfo{volume}{27}, \bibinfo{number}{3} (\bibinfo{year}{1948}), \bibinfo{pages}{379--423}.
\newblock
\urldef\tempurl%
\url{https://doi.org/10.1002/j.1538-7305.1948.tb01338.x}
\showDOI{\tempurl}


\bibitem[Souag et~al\mbox{.}(2018a)]%
        {10.1007/s00766-017-0279-5}
\bibfield{author}{\bibinfo{person}{Amina Souag}, \bibinfo{person}{Ra\'{u}l Mazo}, \bibinfo{person}{Camille Salinesi}, {and} \bibinfo{person}{Isabelle Comyn-Wattiau}.} \bibinfo{year}{2018}\natexlab{a}.
\newblock \showarticletitle{Using the AMAN-DA Method to Generate Security Requirements: A Case Study in the Maritime Domain}.
\newblock \bibinfo{journal}{\emph{Requir. Eng.}} \bibinfo{volume}{23}, \bibinfo{number}{4} (\bibinfo{date}{nov} \bibinfo{year}{2018}), \bibinfo{pages}{557–580}.
\newblock
\showISSN{0947-3602}
\urldef\tempurl%
\url{https://doi.org/10.1007/s00766-017-0279-5}
\showDOI{\tempurl}


\bibitem[Souag et~al\mbox{.}(2018b)]%
        {Souag2018AMANDA}
\bibfield{author}{\bibinfo{person}{Amina Souag}, \bibinfo{person}{Ra{\'u}l Mazo}, \bibinfo{person}{Camille Salinesi}, {and} \bibinfo{person}{Isabelle Comyn-Wattiau}.} \bibinfo{year}{2018}\natexlab{b}.
\newblock \showarticletitle{Using the AMAN-DA method to generate security requirements: a case study in the maritime domain}.
\newblock \bibinfo{journal}{\emph{Requirements Engineering}}  \bibinfo{volume}{23} (\bibinfo{year}{2018}), \bibinfo{pages}{557--580}.
\newblock


\bibitem[Steinmann and Ochoa(2022)]%
        {9736282}
\bibfield{author}{\bibinfo{person}{Jessica Steinmann} {and} \bibinfo{person}{Omar Ochoa}.} \bibinfo{year}{2022}\natexlab{}.
\newblock \showarticletitle{Supporting Security Requirements Engineering through the Development of The Secure Development Ontology}. In \bibinfo{booktitle}{\emph{2022 IEEE 16th International Conference on Semantic Computing (ICSC)}}. \bibinfo{pages}{151--158}.
\newblock
\urldef\tempurl%
\url{https://doi.org/10.1109/ICSC52841.2022.00031}
\showDOI{\tempurl}


\bibitem[{The OWASP® Foundation}(2008)]%
        {ASVS}
\bibfield{author}{\bibinfo{person}{{The OWASP® Foundation}}.} \bibinfo{year}{2008}\natexlab{}.
\newblock \bibinfo{title}{OWASP Application Security Verification Standard (ASVS)}.
\newblock
\newblock
\newblock
\shownote{Accessed August 26, 2024. \url{https://owasp.org/www-project-application-security-verification-standard/}}.


\bibitem[T{\"u}retken et~al\mbox{.}(2004)]%
        {turetken2004automating}
\bibfield{author}{\bibinfo{person}{Oktay T{\"u}retken}, \bibinfo{person}{Onur Su}, {and} \bibinfo{person}{Onur Demir{\"o}rs}.} \bibinfo{year}{2004}\natexlab{}.
\newblock \showarticletitle{Automating software requirements generation from business process models}. In \bibinfo{booktitle}{\emph{1st Conf. on the Principles of Software Eng.(PRISE’04), Buenos Aires, Argentina}}.
\newblock


\bibitem[Vali and Modiri({[n.\,d.]})]%
        {valiiso}
\bibfield{author}{\bibinfo{person}{Nasser Vali} {and} \bibinfo{person}{Nasser Modiri}.} \bibinfo{year}{[n.\,d.]}\natexlab{}.
\newblock \showarticletitle{ISO/IEC 15408}.
\newblock  (\bibinfo{year}{[n.\,d.]}).
\newblock


\bibitem[Van~Lamsweerde(2004)]%
        {van2004goal}
\bibfield{author}{\bibinfo{person}{Axel Van~Lamsweerde}.} \bibinfo{year}{2004}\natexlab{}.
\newblock \showarticletitle{Goal-oriented requirements enginering: a roundtrip from research to practice [enginering read engineering]}. In \bibinfo{booktitle}{\emph{Proceedings. 12th IEEE International Requirements Engineering Conference, 2004.}} IEEE, \bibinfo{pages}{4--7}.
\newblock


\bibitem[van Lamsweerde and Willemet(1998)]%
        {DBLP:journals/tse/LamsweerdeW98}
\bibfield{author}{\bibinfo{person}{Axel van Lamsweerde} {and} \bibinfo{person}{Laurent Willemet}.} \bibinfo{year}{1998}\natexlab{}.
\newblock \showarticletitle{Inferring Declarative Requirements Specifications from Operational Scenarios}.
\newblock \bibinfo{journal}{\emph{{IEEE} Trans. Software Eng.}} \bibinfo{volume}{24}, \bibinfo{number}{12} (\bibinfo{year}{1998}), \bibinfo{pages}{1089--1114}.
\newblock


\bibitem[Welch(1947)]%
        {1947The}
\bibfield{author}{\bibinfo{person}{B~L Welch}.} \bibinfo{year}{1947}\natexlab{}.
\newblock \showarticletitle{The generalization of Student's problem when several different population variances are involved}.
\newblock \bibinfo{journal}{\emph{Biometrika}} \bibinfo{volume}{34}, \bibinfo{number}{1-2} (\bibinfo{year}{1947}), \bibinfo{pages}{28--35}.
\newblock


\bibitem[Ye et~al\mbox{.}(2023)]%
        {ye2023comprehensive}
\bibfield{author}{\bibinfo{person}{Junjie Ye}, \bibinfo{person}{Xuanting Chen}, \bibinfo{person}{Nuo Xu}, \bibinfo{person}{Can Zu}, \bibinfo{person}{Zekai Shao}, \bibinfo{person}{Shichun Liu}, \bibinfo{person}{Yuhan Cui}, \bibinfo{person}{Zeyang Zhou}, \bibinfo{person}{Chao Gong}, \bibinfo{person}{Yang Shen}, {et~al\mbox{.}}} \bibinfo{year}{2023}\natexlab{}.
\newblock \showarticletitle{A comprehensive capability analysis of gpt-3 and gpt-3.5 series models}.
\newblock \bibinfo{journal}{\emph{arXiv preprint arXiv:2303.10420}} (\bibinfo{year}{2023}).
\newblock


\bibitem[Yu et~al\mbox{.}(1995)]%
        {DBLP:conf/coopis/YuBDM95}
\bibfield{author}{\bibinfo{person}{Eric S.~K. Yu}, \bibinfo{person}{Philippe~Du Bois}, \bibinfo{person}{Eric Dubois}, {and} \bibinfo{person}{John Mylopoulos}.} \bibinfo{year}{1995}\natexlab{}.
\newblock \showarticletitle{From Organization Models to System Requirements: {A} 'Cooperating Agents' Approach}. In \bibinfo{booktitle}{\emph{Proceedings of the Third International Conference on Cooperative Information Systems (CoopIS-95), May 9-12}}. \bibinfo{pages}{194--204}.
\newblock


\bibitem[Zhao et~al\mbox{.}(2024a)]%
        {zhao2024enhancingllmbasedautomatedprogram}
\bibfield{author}{\bibinfo{person}{Jiuang Zhao}, \bibinfo{person}{Donghao Yang}, \bibinfo{person}{Li Zhang}, \bibinfo{person}{Xiaoli Lian}, {and} \bibinfo{person}{Zitian Yang}.} \bibinfo{year}{2024}\natexlab{a}.
\newblock \bibinfo{title}{Enhancing LLM-Based Automated Program Repair with Design Rationales}.
\newblock
\newblock
\showeprint[arxiv]{2408.12056}~[cs.SE]
\urldef\tempurl%
\url{https://arxiv.org/abs/2408.12056}
\showURL{%
\tempurl}


\bibitem[Zhao et~al\mbox{.}(2024b)]%
        {zhao2024novelapproachautomateddesign}
\bibfield{author}{\bibinfo{person}{Jiuang Zhao}, \bibinfo{person}{Zitian Yang}, \bibinfo{person}{Li Zhang}, \bibinfo{person}{Xiaoli Lian}, {and} \bibinfo{person}{Donghao Yang}.} \bibinfo{year}{2024}\natexlab{b}.
\newblock \bibinfo{title}{A Novel Approach for Automated Design Information Mining from Issue Logs}.
\newblock
\newblock
\showeprint[arxiv]{2405.19623}~[cs.SE]
\urldef\tempurl%
\url{https://arxiv.org/abs/2405.19623}
\showURL{%
\tempurl}


\bibitem[Zhao et~al\mbox{.}(2023)]%
        {ReqGen}
\bibfield{author}{\bibinfo{person}{Ziyan Zhao}, \bibinfo{person}{Li Zhang}, \bibinfo{person}{Xiaoli Lian}, \bibinfo{person}{Xiaoyun Gao}, \bibinfo{person}{Heyang Lv}, {and} \bibinfo{person}{Lin Shi}.} \bibinfo{year}{2023}\natexlab{}.
\newblock \showarticletitle{ReqGen: Keywords-Driven Software Requirements Generation}.
\newblock \bibinfo{journal}{\emph{Mathematics}}  \bibinfo{volume}{11} (\bibinfo{date}{01} \bibinfo{year}{2023}), \bibinfo{pages}{332}.
\newblock
\urldef\tempurl%
\url{https://doi.org/10.3390/math11020332}
\showDOI{\tempurl}


\bibitem[Zhu et~al\mbox{.}(2018)]%
        {zhu2018texygen}
\bibfield{author}{\bibinfo{person}{Yaoming Zhu}, \bibinfo{person}{Sidi Lu}, \bibinfo{person}{Lei Zheng}, \bibinfo{person}{Jiaxian Guo}, \bibinfo{person}{Weinan Zhang}, \bibinfo{person}{Jun Wang}, {and} \bibinfo{person}{Yong Yu}.} \bibinfo{year}{2018}\natexlab{}.
\newblock \showarticletitle{Texygen: A benchmarking platform for text generation models}. In \bibinfo{booktitle}{\emph{The 41st international ACM SIGIR conference on research \& development in information retrieval}}. \bibinfo{pages}{1097--1100}.
\newblock


\bibitem[Łukasiewicz and Cygańska(2019)]%
        {8860028}
\bibfield{author}{\bibinfo{person}{Katarzyna Łukasiewicz} {and} \bibinfo{person}{Sara Cygańska}.} \bibinfo{year}{2019}\natexlab{}.
\newblock \showarticletitle{Security-oriented agile approach with AgileSafe and OWASP ASVS}. In \bibinfo{booktitle}{\emph{2019 Federated Conference on Computer Science and Information Systems (FedCSIS)}}. \bibinfo{pages}{875--878}.
\newblock
\urldef\tempurl%
\url{https://doi.org/10.15439/2019F213}
\showDOI{\tempurl}


\end{thebibliography}

\end{document}